\documentclass[11pt,psfig]{article}
\usepackage{a4wide}
\usepackage{graphicx}
\usepackage{subfigure}
\usepackage[centertags]{amsmath}
\usepackage{amssymb}
\usepackage{dsfont}
\usepackage{cite}
\newcommand{\unity}{{ 1\:\!\!\!\mbox{I}}}

\newcommand{\Z}{{\mathbb Z}}
\newcommand{\R}{\mathcal R}
\newcommand{\OR}{\Omega {\mathcal R}}

\newcommand{\au}{\mathcal{A}}
\newcommand{\msu}{\mathcal{M}}
\newcommand{\lu}{\mathcal{L}}

\newcommand{\3}{{\bf 3}}
\newcommand{\2}{{\bf 2}}

\newcommand{\ov}{\overline}
\newcommand{\odd}{\text{odd}}
\newcommand{\even}{\text{even}}

\newcommand{\eps}{\varepsilon_}
\newcommand{\Teps}{\Tilde{\varepsilon}_}

\newcommand{\barre}[1]{%
        \setbox1=\hbox{$#1$} \dimen2=\ht1 \dimen3=\dp1 \dimen4=\wd1
        \setbox2=\hbox{\sl /}
        \dimen1=\wd1 \advance\dimen1 by -\wd2 \divide\dimen1 by 2
        \advance\dimen1 by \wd2 \advance\dimen1 by 0.4pt
        \setbox3=\hbox to \wd1{\hss \box1 \kern -\dimen1 \box2\hss}
        \ht3=\dimen2 \dp3=\dimen3 \wd3=\dimen4
        \box3
        }

\begin{document}
\pagestyle{empty}
\rightline{KUL-TF-04/13}
\rightline{FT-UAM-04-05}
\rightline{IFT-UAM/CSIC-04-12}
\begin{flushright}
                     hep-th/0404055
\end{flushright}
\vskip 1cm

\begin{center}
{\huge Getting just the Supersymmetric Standard Model\\}
{\huge at Intersecting Branes on the $\mathbb{Z}_6$-orientifold}
\vspace*{5mm} \vspace*{7mm}
\end{center}
\vspace*{5mm} \noindent
\vskip 0.5cm
\centerline{\bf Gabriele Honecker$^1$ and Tassilo Ott$^2$}
\vskip 1cm
\centerline{\em $^1$  Departamento de F\'{\i}sica Te\'{o}rica C-XI and
Instituto de F\'{\i}sica Te\'{o}rica C-XVI,}
\centerline{\em Universidad Aut\'{o}noma de Madrid, Cantoblanco, 28049 Madrid, Spain}
\centerline{\tt E-mail: gabriele@th.physik.uni-bonn.de}
\vskip .3cm
\centerline{\em $^2$ Institute for Theoretical Physics, KULeuven}
\centerline{\em Celestijnenlaan 200 D, 3001 Leuven, Belgium}
\centerline{\tt E-mail: tassilo.ott@fys.kuleuven.ac.be}

\vskip1cm

\centerline{\bf Abstract} \noindent
In this paper, globally $\mathcal{N}=1$
supersymmetric configurations of intersecting D6-branes on the
$\mathbb{Z}_6$ orientifold are discussed, involving also
fractional branes. It turns out rather miraculously that one is
led almost automatically to just \textit{one} particular class of
5 stack models containing the SM gauge group, which all have the
same chiral spectrum. The
further discussion shows that these models can be understood as
exactly the supersymmetric standard model without any exotic
chiral symmetric/antisymmetric matter. The superpartner of the
Higgs finds a natural explanation and the hypercharge remains
massless. However, the non-chiral spectrum within the model class
is very different and does not in all cases allow for  a
$\mathcal{N}=2$ low energy field theoretical understanding of the
necessary breaking $U(1)\times U(1)\rightarrow U(1)$ along the
Higgs branch, which is needed in order to get the standard Yukawa
couplings.\\
Also the left-right symmetric models belong to exactly {\it one} class of
chiral spectra, where the two kinds of exotic chiral fields can have the
interpretation of forming a composite Higgs.\\
The aesthetical beauty of these models, involving only
non-vanishing intersection numbers of an absolute value three,
seems to be unescapable.

\vskip .3cm


\newpage

\setcounter{page}{1} \pagestyle{plain}

\section{Introduction} \label{intro}
String Theory claims to be the correct unifying theory of gravity
and elementary particle physics. As the latter, it should contain
the standard model as a low energy limit, which is a chiral gauge
theory. Since it was realized that chiral matter is possible in
the context of intersecting D-branes \cite{Berkooz:1996aa}, many
different approaches have been taken, the most successful ones
being constructions with intersecting D6-branes in type IIA
orientifolds \cite{Sagnotti:1987tw, Blumenhagen:1999db,
Blumenhagen:1999ev, Blumenhagen:2000wh, Blumenhagen:2000ea,
Angelantonj:2000hi,Aldazabal:2000cn, Aldazabal:2000dg, Forste:2000hx,
Forste:2001gb, Kachru:1999vj, Uranga:2003pz} (for recent reviews
of the topic see \cite{Ott:2003yv,MarchesanoBuznego:2003hp, Gorlich:2004zs,
Angelantonj:2002ct}). The branes wrap special Lagrangian
3-cycles of the compact space in these models, which in general
can be a simple toroidal, orbifolded or Calabi-Yau space, where
the worldsheet parity symmetry $\Omega$ together with a space time
symmetry $\R$ is modded out. In this picture, chiral fermions are
localized at the intersections of the different stacks of
D6-branes and consequently are 4-dimensional, whereas the gauge
fields live on the whole worldvolume of the branes and by this are
7-dimensional. Finally, gravity is contained in the closed string
sector of the theory and lives in the 10-dimensional bulk.

The massless chiral fermion spectrum can be determined in all
three cases, as it only depends on the homology of the 3-cycles
\cite{Blumenhagen:2002wn, Blumenhagen:2002vp}. On the other hand,
the non-chiral part of the spectrum depends on the closed string
moduli and requires the worldsheet CFT computation of the one loop
amplitude that can be explicitly obtained only in the case of the
torus or orbifolded torus. The 6-torus usually is assumed to be
factorized in three 2-tori, i.e. $T^6=T^2\times T^2 \times T^2$.
Every 3-cycle factorizes into three 1-cycles, one
on every torus, and one obtains a nice geometrical picture.

In such an approach, a model containing only the standard model
matter in the chiral sector has been obtained \cite{Ibanez:2001nd}
(for similar subsequent models see \cite{Kokorelis:2002ip,
Kokorelis:2002zz}). But shortly afterwards, it was realized that
this model is unstable due to the uncancelled NS-NS-tadpoles
becoming manifest in the runaway behavior of the complex structure
and dilaton moduli \cite{Blumenhagen:2001te, Blumenhagen:2001mb}
(see also the remarks in \cite{Rabadan:2001mt}). The first instability can be
cured by making the transition to a $\Z_3$-orientifold (where the
exact realization of the standard model is different), but the
dilaton instability cannot. 
Although this instability
might be interesting from the perspective of cosmology and
particularly inflation at first sight, it has been found, that
only under very special and rather unsatisfactory requirements
(for instance some moduli have to be fixed by an unknown
mechanism), the remaining modulus could act as the inflaton
\cite{Blumenhagen:2002ua} (for a different approach see e.g.
\cite{Garcia-Bellido:2001ky}). Beside that fact, inflation has to
end and today only a very small cosmological constant is observed,
which would require a very unnatural fine-tuning to comply with a
natural timescale of for instance the dilaton instability.

Therefore, in the recent past another road has mainly been taken,
namely the attempt to construct instead a ${\cal N}=1$
supersymmetric standard model (or supersymmetric $SU(5)$ GUTs and
Pati-Salam models respectively). A first supersymmetric
three-generation standard-like model has been constructed in
\cite{Cvetic:2001tj, Cvetic:2001nr}. Nevertheless, the goal so far
has been achieved only with moderate success: either the
constructions are plagued with a large amount of exotic chiral
matter, as in the $\Z_2\times \Z_2$ orientifold
of~\cite{Cvetic:2001tj, Cvetic:2001nr, Cvetic:2002pj,
Cvetic:2003xs, Cvetic:2004ui, Larosa:2003mz}, which might be cured
by a confinement of the exotic chiral matter into composite fields
coming from the strong infrared dynamics of the hidden sector
\cite{Cvetic:2002qa, Cvetic:2002wh}. In the other cases, brane
recombinations of non-Abelian gauge groups are needed, as in the
$\Z_4$ orientifold of \cite{Blumenhagen:2002gw} or the $\Z_4\times
\Z_2$ \cite{Honecker:2003vq, Honecker:2003vw}, giving rise to
Pati-Salam-Models, and leading to non-flat and non-factorizable
branes which give up the complete predictability of the worldsheet
CFT approach. Another possibility are the constructions
of~\cite{Cremades:2002te, Cremades:2002qm}, where only the sectors
between certain branes are locally ${\cal N}=1$-supersymmetric
(Q-SUSY theories), but the setting as a whole is not, meaning that
the NS-NS-tadpole is not cancelled. For similar constructions see
\cite{Kokorelis:2002iz, Kokorelis:2003jr}.

If a realistic globally ${\cal N}=1$ supersymmetric standard model
with the right chiral spectrum was found, many phenomenological
properties could be discussed \cite{Witten:2002wb}, such as for
instance proton decay \cite{Klebanov:2003my} or the running of the
gauge couplings \cite{Cvetic:2002qa, Lust:2003ky} and a possible
gauge unification \cite{Blumenhagen:2003jy}, the generation of
masses \cite{Cremades:2002va, Kors:2004dx, Kitazawa:2004hz}, and
the precise realization of the Higgs mechanism(s)
\cite{Cremades:2002cs} and Yukawa couplings \cite{Cremades:2003qj,
Cvetic:2003ch, Abel:2003vv}, leading towards the goal of making
contact with experimental reality \cite{Cremades:2002va}. It could
even lead to an understanding of supersymmetry breaking
\cite{Cvetic:2003yd, Kors:2003wf}.

Such a model can be compared to the minimal supersymmetric
standard model (MSSM), which is claimed to be the best candidate
for the search for supersymmetry at the LHC (for a good review see
for instance \cite{Louis:1998rx}). This model has more predictive
properties than a general supersymmetric standard model, it
contains only a minimal Higgs sector of two $SU(2)$ doublets and
their superpartners whose hypercharge is exactly opposite, the
declaration that all scalar masses are the same, all gaugino
masses are the same, some statements about soft supersymmetry
breaking (in the context of intersecting branes see
\cite{Kors:2003wf}). 
In the past, the term MSSM often has
been used very sloppy within string model building, basically
meaning only the correct chiral spectrum. For a recent overview
about the actual status of D6-brane constructions see
\cite{Lust:2004ks, Lust:2004wr}.

A new possibility that has been explored is to get
phenomenologically interesting models in a completely different
corner of Calabi-Yau moduli space, namely at the Gepner points
\cite{Blumenhagen:2003su, Blumenhagen:2004cg, Blumenhagen:2004qu,
Aldazabal:2003ub, Aldazabal:2004by}. Even chiral supersymmetric models
with the standard model spectrum have been obtained already in
this corner of moduli space \cite{Dijkstra:2004ym}(for an
introduction see \cite{Huiszoon:2002uj}).

In this paper, the aim of getting a precise realization of a
supersymmetric standard model will be continued furthermore on the
compact space $T^6/\Z_6$, which again corresponds to a ${\cal
N}=2$ background of type II theory, following the classification
of orbifolds in~\cite{Dixon:1985jw, Dixon:1986jc}. For this
background (which geometrically can be defined in two different
consistent ways on the  2-tori $T^2_k$), the $\Z_2$-twisted sector
in both cases contributes a non-vanishing number $h_{2,1}$ to the
number of complex structure deformations and so contains twisted
3-cycles, requiring the introduction of fractional branes
\cite{Diaconescu:1998br, Diaconescu:1999dt}, similar to the
$\Z_4$-orientifold of~\cite{Blumenhagen:2002gw}.

The organization of the paper is as follows. The geometry of the
orbifold is  discussed in detail in section
\ref{Sec:geometry}, including the definition of bulk and
fractional cycles and an integral basis of homology, the
orientifold plane and the resulting R-R-tadpole conditions for all
possible choices of {\bf A}- and {\bf B}-tori\footnote{They correspond to
a vanishing or non-vanishing NS-NS 2-form flux $b$ in the T-dual
F-flux picture of D9-branes.}. Finally, the conditions for ${\cal N}=1$
supersymmetry on the bulk and exceptional cycles are derived.

Section \ref{Sec:spectrum} discusses the calculation of the
one-loop amplitudes and the transformation to the tree-channel,
preparing the calculation of the chiral and non-chiral open string
spectrum as well as clarifying the connection between the
computation of cycles and string loop amplitudes.

Subsequently, there is a detailed discussion on anomalies and
the generalized Green-Schwarz-mechanism in section
\ref{Sec:Anomalies_GS}.

In section \ref{Sec:SUSYmodels}, we finally come to the systematic
search of phenomenologically interesting models with a different
number of stacks, leading to a detailed presentation of the
explicit supersymmetric standard model which we have found in
section~\ref{Sec:SUSY-SM}. There is also a short discussion on
another possible left/right symmetric model in
section~\ref{Sec:LR-sym-SUSY-models}.

The conclusions and prospects are given in chapter
\ref{Sec:Conclusions}.

\noindent
Some technical details are collected in
appendices~\ref{AppSec:BasisTori} to~\ref{AppSec:Z6Z3}.

\section{Geometry of the $T^6/\Z_6$ orbifold}
\label{Sec:geometry}
In this section, the geometric setting of the
six dimensional compact space including D6-branes, O6-planes,
cancellation of RR tadpoles and supersymmetry are discussed.

If
one assumes a factorization of the $T^6$ into three 2-tori, then
the $T^6/\Z_6$ orbifold is generated by a rotation $\theta$ of the
form
\begin{equation}
\theta:\qquad z^k\rightarrow e^{2\pi i v_k}z^k \qquad
\text{for}\ \ k=1,2,3
\end{equation}
where the shift vector $v$ is given by \mbox{$v=(1/6,1/6,-1/3)$},
see for instance \cite{Dixon:1985jw, Dixon:1986jc}, and $z^k$ is
the complex coordinate on $T^2_k$.\footnote{There exists a second,
inequivalent possibility with $\mathbb{Z}_6$ symmetry, having the
shift vector \mbox{$v=(1/6,1/3,-1/2)$}, which is often denoted as
$\mathbb{Z}'_6$. Both symmetric orbifolds in the IIA picture are
T-dual to asymmetric orbifolds in the IIB background. Our models
are therefore not T-dual to the symmetric $\Z_6$ orbifold of IIB
in~\cite{Cvetic:2000st}.}

\begin{figure}[t]
\begin{center}
\includegraphics[scale=0.8]{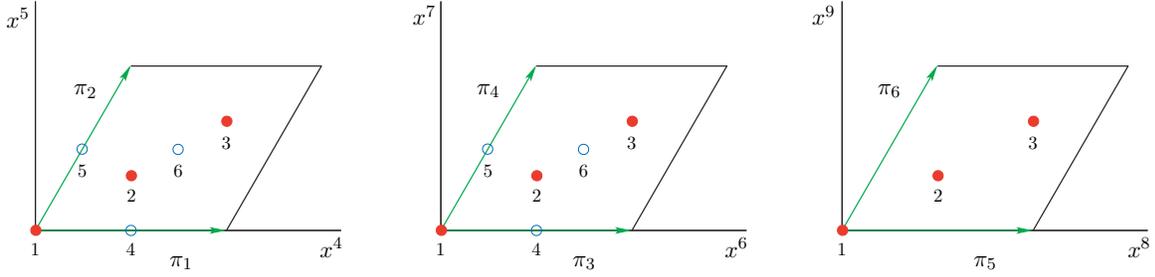}
\end{center}
\caption{Fixed points of the $T^6/\Z_6$ orbifold. Full circles
denote
  $\theta^2$ fixed points on $T^2_1 \times T^2_2$, empty circles additional
  $\theta^3$ fixed points. On $T^2_3$, the points 1,2,3 are fixed under
  $\theta$, the whole $T^2_3$ is fixed under $\theta^3$.
The coordinates are depicted for the {\bf AAA} torus. The details of the choices of complex structures are given in
section \ref{Subsec:OProjections} and appendix \ref{AppSec:BasisTori}.} \label{Fig:Z6torifixedpoints}
\end{figure}

Apart from the $\Z_6$ fixed points, additional points are fixed
under the $\Z_3$ subsymmetry generated by $\theta^2$. At all these
fixed points, exceptional 2-cycles of zero volume\footnote{The
volume is not zero in the stringy sense, i.e. a non-vanishing
string tension is generated by a discrete NS-NS two-form
background in the ten dimensional language. For more details see
e.g. the review article~\cite{Bertolini:2003iv}.} are stuck. Under
the $\Z_2$ subsymmetry generated by $\theta^3$, the whole third
2-torus $T^2_3$ is fixed, and exceptional 3-cycles which are
products of 2-cycles stuck at the $\Z_2$ fixed points on $T^2_1
\times T^2_2$ times 1-cycles on the fix-torus $T^2_3$ appear. The
geometry is depicted in figure~\ref{Fig:Z6torifixedpoints}, where
a specific complex structure of the torus from six possibilities
has been chosen. This is explained in more detail in
\ref{Subsec:OProjections} and appendix \ref{AppSec:BasisTori}.

The Hodge numbers of the $\Z_6$ orbifold are as follows (see
e.g.~\cite{Blumenhagen:1999ev} for the closed string spectrum,
~\cite{Aldazabal:1998mr} for the number of untwisted and twisted
moduli and~\cite{Klein:2000hf} for the Hodge numbers explicitly),
\begin{equation}
\begin{aligned}\label{Eq:HodgeNumbers}
h^U_{1,1}&=5, \qquad h^{\theta}_{1,1}=3, \qquad
h^{\theta^2}_{1,1}=15,
\qquad h^{\theta^3}_{1,1}=6,\\
h^U_{2,1}&=0, \qquad h^{\theta}_{2,1}=0, \qquad
h^{\theta^2}_{2,1}=0, \qquad h^{\theta^3}_{2,1}=5.
\end{aligned}
\end{equation}
In the following, we will only consider the 3-cycles, but not the
2-cycles, since only intersections of D6-branes wrapping different
3-cycles in the compact dimensions give rise to
chiral fermions.\\
According to the value of the third Betti number,
$b_3=2+2h_{2,1}$, two  independent `bulk' 3-cycles are inherited
from the six-torus and ten additional exceptional 3-cycles arise
at the $\Z_2$ fix-points.

\subsection{Bulk 3-cycles}\label{Subsec:bulk3cycles}

Any basic factorizable 3-cycle on $\prod_{k=1}^3 T^2_k$ can be
represented in terms of the basic 1-cycles of each 2-torus,
$\pi_{2k-1}$ and  $\pi_{2k}$, as a direct product $\pi_{i,j,m}
=\pi_i \otimes \pi_j \otimes \pi_m$. Out of the $2^3$ different
combinations, one can construct only two linearly independent bulk
cycles which are invariant under the orbifold action. A convenient
choice for them is given by
\begin{equation}
\begin{aligned}\label{Eq:bulkcycles}
\rho_1 = 2\left[(1+\theta +\theta^2)\pi_{1,3,5}\right]
&=2\left(\pi_{1,3,5}+\pi_{2,4,-6}+\pi_{2-1,4-3,6-5}\right)\\
 &= 2\left( \pi_{1,4,5}+\pi_{1,3,6}+\pi_{2,3,5}-\pi_{1,4,6}-\pi_{2,4,5}-\pi_{2,3,6} \right),\\
\rho_2 = 2\left[(1+\theta +\theta^2)\pi_{2,3,5}\right]
&=2\left(\pi_{2,3,5}+\pi_{2-1,4,-6}+\pi_{-1,4-3,6-5} \right)\\
 &= 2\left(\pi_{1,4,5}+\pi_{1,3,6}+\pi_{2,3,5}-\pi_{1,3,5}-\pi_{2,4,6}\right).
\end{aligned}
\end{equation}
The factor of two in~(\ref{Eq:bulkcycles}) arises due to the
trivial action of $\theta^3$ on any 3-cycle. Any orbifold
invariant non-factorizable 3-cycle can be written as a linear
combination of these two bulk cycles.

The coefficients of the factorizable 3-cycles are determined by
the wrapping numbers $n_k$ and $m_k$ along the basic 1-cycles
$\pi_{2k-1}$ and $\pi_{2k}$ on the 2-torus $T^2_k$ and their
orbifold images,
\begin{equation}\label{Eq:OrbitWrappingNumbers}
\left(\begin{array}{cc} n_1 & m_1 \\ n_2 & m_2 \\n_3 & m_3
\end{array}\right)\stackrel{\theta}{\longrightarrow}
\left(\begin{array}{cc} -m_1 & n_1+m_1 \\ -m_2 & n_2+m_2 \\m_3 &
-(n_3+m_3)
\end{array}\right)\stackrel{\theta}{\longrightarrow}
\left(\begin{array}{cc} -(n_1+m_1) & n_1 \\ -(n_2+m_2) & n_2
\\-(n_3+m_3) & n_3
\end{array}\right).
\end{equation}
Starting with the 3-cycle $(n^a_1 \pi_1 +  m^a_1 \pi_2) \otimes
(n^a_2 \pi_3 +  m^a_2 \pi_4) \otimes (n^a_3 \pi_5 +  m^a_3 \pi_6)$
and adding its orbifold images, the invariant bulk 3-cycle is of
the form
\begin{equation}\label{Eq:Shortcycles}
\Pi_a=Y_a \rho_1 + Z_a \rho_2,
\end{equation}
where the coefficients $Y_a$ and $Z_a$ are given by linear
combinations of products of the elementary wrapping numbers,
\begin{equation}\label{Eq:bulkCoeffYZ}
\begin{aligned}
Y_a & \equiv n^a_1 n^a_2 n^a_3-\sum_{i \neq j \neq k \neq i} m^a_i m^a_j n^a_k -  m^a_1 m^a_2 m^a_3,\\
Z_a  & \equiv \sum_{i \neq j \neq k \neq i} m^a_i m^a_j n^a_k +
\sum_{i \neq j \neq k \neq i} m^a_i n^a_j n^a_k.
\end{aligned}
\end{equation}

In computing the intersection numbers of the two independent bulk
cycles which are invariant under an $\Z_N$ action, the orbifold
projection has to be taken into account,
\begin{equation}\label{Eq:OrbIntersection}
\Pi_a \circ \Pi_b =\frac{1}{N}\left(\sum_{i=0}^{N-1}\theta^i \pi_a
\right) \circ\left(\sum_{j=0}^{N-1}\theta^j \pi_b \right).
\end{equation}
This leads to the intersection numbers of the two fundamental bulk
cycles~(\ref{Eq:bulkcycles}),
\begin{equation}\label{Eq:fundbulkIntersection}
\rho_1 \circ \rho_2 =-2, \qquad \rho_1 \circ \rho_1=\rho_2 \circ
\rho_2=0,
\end{equation}
and for general bulk cycles $\Pi_a=Y_a \rho_1 + Z_a \rho_2$, we
always obtain even intersection numbers,
\begin{equation}\label{Eq:OrbitbulkIntersection}
I_{ab}\equiv \Pi_a \circ \Pi_b = 2(Z_a Y_b - Y_a Z_b).
\end{equation}


\subsection{Exceptional 3-cycles}\label{Subsec:ex3cycles}

Exceptional 3-cycles in the orbifold limit only occur in the
$\theta^3$ sector. They consist of products of 2-cycles which are
stuck at the $\Z_2$ fixed points and have zero volume in the
orbifold limit on $T^2_1 \times T^2_2$ times a 1-cycle on $T^2_3$.
The fixed point $1$ in the origin of each 2-torus is already
invariant under the orbifold generator $\theta$. The other three
$\Z_2$ fixed points are permuted by the $\Z_6$ symmetry in the
following way,
\begin{equation}\label{Eq:thetaonfixedpoints}
\theta (4)=5, \qquad \theta (5)=6, \qquad \theta (6)=4.
\end{equation}
Similarly, every 1-cycle on $T^2_3$ is invariant under $\theta^3$,
but the $\Z_6$ generator acts as follows,
\begin{equation}
\theta(\pi_5)=-\pi_6, \qquad \theta(\pi_6)=\pi_{5-6}=\pi_5- \pi_6.
\end{equation}
Thus, the exceptional 3-cycles are given by orbifold invariant
combinations of the products of 2- and 1-cycles,
\begin{equation}
(1+\theta+\theta^2)\left(e_{ij} \otimes \pi_k \right) \qquad
\text{ with } \qquad i,j=1,4,5,6; \; k=5,6,
\end{equation}
where the orbifold images are given by $\theta(e_{ij} \otimes
\pi_k )=e_{\theta(i)\theta(j)}\otimes \theta(\pi_k)$. This ansatz
leads to ten linearly independent exceptional cycles,
\begin{equation}\label{Eq:exceptionalcycles}
\begin{aligned}
\varepsilon_1 &= \left( e_{41}-e_{61}\right) \otimes \pi_5 +\left(
e_{61}-e_{51}\right) \otimes \pi_6, \qquad \Tilde{\varepsilon}_1 =
\left( e_{51}-e_{61}\right) \otimes \pi_5 +\left(
e_{41}-e_{51}\right)
\otimes \pi_6,\\
\varepsilon_2 &= \left( e_{14}-e_{16}\right) \otimes \pi_5 +\left(
e_{16}-e_{15}\right) \otimes \pi_6, \qquad \Tilde{\varepsilon}_2 =
\left( e_{15}-e_{16}\right) \otimes \pi_5 +\left(
e_{14}-e_{15}\right)
\otimes \pi_6,\\
\varepsilon_3 &= \left( e_{44}-e_{66}\right) \otimes \pi_5 +\left(
e_{66}-e_{55}\right) \otimes \pi_6, \qquad \Tilde{\varepsilon}_3 =
\left( e_{55}-e_{66}\right) \otimes \pi_5 +\left(
e_{44}-e_{55}\right)
\otimes \pi_6,\\
\varepsilon_4 &= \left( e_{45}-e_{64}\right) \otimes \pi_5 +\left(
e_{64}-e_{56}\right) \otimes \pi_6, \qquad \Tilde{\varepsilon}_4 =
\left( e_{56}-e_{64}\right) \otimes \pi_5 +\left(
e_{45}-e_{56}\right)
\otimes \pi_6,\\
\varepsilon_5 &= \left( e_{46}-e_{65}\right) \otimes \pi_5 +\left(
e_{65}-e_{54}\right) \otimes \pi_6, \qquad \Tilde{\varepsilon}_5 =
\left( e_{54}-e_{65}\right) \otimes \pi_5 +\left(
e_{46}-e_{54}\right) \otimes \pi_6.
\end{aligned}
\end{equation}

This construction resembles very much the one in
\cite{Blumenhagen:2002gw}. The intersection numbers of exceptional
cycles are computed from the self intersection number $-2$ of any
exceptional cycle $e_{ij}$ stuck at a $\Z_2$ singularity and the
intersection numbers of the 1-cycles on $T^2_3$, taking into
account the factors from the orbifold projection similarly to
equation~(\ref{Eq:OrbIntersection}). The  result
\begin{equation}
\varepsilon_i \circ \Tilde{\varepsilon}_j =-2\delta_{ij}, \qquad
\varepsilon_i \circ \varepsilon_j  =\Tilde{\varepsilon}_i \circ
\Tilde{\varepsilon}_j=0,
\end{equation}
leads to the intersection matrix for exceptional cycles
\begin{equation}\label{Eq:IntersectionMatrixExcycles}
I_{\varepsilon}=\bigoplus_{j=1}^5 \left(\begin{array}{cc} 0 & -2
\\ 2 & 0
\end{array}\right).
\end{equation}


\subsection{An integral basis}\label{Subsec:basis}

The intersection numbers of pure bulk
cycles~(\ref{Eq:fundbulkIntersection}) and pure exceptional
cycles~(\ref{Eq:IntersectionMatrixExcycles}) are always even. It
is therefore possible to construct fractional cycles of the form
$\frac{1}{2}\Pi^{bulk}+\frac{1}{2}\Pi^{exceptional}$ which form an
unimodular lattice as required by Poincar\'e
duality~\cite{GriffithHarris:2000qw}.

For example, a D6-brane with wrapping numbers
$(n_1,m_1;n_2,m_2;n_3,m_3)=(1,0;1,0;1,0)$ and its orbifold images
wrap the bulk cycle $\rho_1$ and pass through the fixed points
$e_{kl}$ with $k,l \in \{1,4\}$ as well as their orbifold images.
The orbits of the three non-trivial fixed points $e_{14}, e_{41},
e_{44}$ generate the exceptional cycles $\varepsilon_1$,
$\varepsilon_2$ and $\varepsilon_3$ while the orbit of $e_{11}$
vanishes. In this case, a valid fractional cycle is given by
\begin{equation}
\frac{1}{2}\rho_1 \pm \frac{1}{2}\left(\varepsilon_1 \pm
\varepsilon_2 \pm \varepsilon_3 \right)
\end{equation}
with arbitrary relative signs for the exceptional contributions.

This argument can be repeated for different wrapping numbers and
3-cycles which do not pass through $e_{11}$ but instead are
displaced by $\sum_{i=1}^4 \sigma_i \pi_i$ (with $\sigma_i \in
\{0,\frac{1}{2}\}$) from the origins of $T^2_1$ and $T^2_2$. All
possible combinations of wrapping numbers of the bulk 3-cycles
with fixed points are listed in table~\ref{AppTab:WrapFPZ6}. From
this, we can obtain a basis for the unimodular lattice of 3-cycles
\begin{equation}
 \begin{aligned}
& 
\alpha_1  =\frac{1}{2}\rho_1 +\frac{1}{2}\left( \varepsilon_1
+\varepsilon_2 +\varepsilon_3\right),
\qquad 
\alpha_5  =\frac{1}{2}\left( \varepsilon_1 - \varepsilon_3 +
\varepsilon_4 + \varepsilon_5 \right),
\\
& \alpha_2  =\frac{1}{2}\rho_2 -\frac{1}{2}\left(
\Tilde{\varepsilon}_1 + \Tilde{\varepsilon}_2+
\Tilde{\varepsilon}_3 \right),
\qquad 
\alpha_6  =\frac{1}{2}\left( -\Tilde{\varepsilon}_1 +
\Tilde{\varepsilon}_3 + \Tilde{\varepsilon_4} -
\Tilde{\varepsilon}_5 \right),
\\
& \alpha_3  =\frac{1}{2}\rho_1 +\frac{1}{2}\left( \varepsilon_1 -
\varepsilon_2 +\varepsilon_3\right),
\qquad 
\alpha_7  =\frac{1}{2}\left( \Tilde{\varepsilon}_1 -
\Tilde{\varepsilon}_3 - \Tilde{\varepsilon}_4 -
\Tilde{\varepsilon}_5 \right),
\\
& \alpha_4  =\frac{1}{2}\rho_2 -\frac{1}{2}\left(
\Tilde{\varepsilon}_1 - \Tilde{\varepsilon}_2+
\Tilde{\varepsilon}_3 \right),
\qquad 
\alpha_8  =\frac{1}{2}\left( \varepsilon_1 - \varepsilon_3 +
\varepsilon_4 - \varepsilon_5 \right),
\\
& \alpha_9  = -\frac{1}{2}\rho_1 +\frac{1}{2}\left( -\varepsilon_3
+ \varepsilon_4-\Tilde{\varepsilon}_1
+\Tilde{\varepsilon}_3+ 2 \Tilde{\varepsilon}_4 \right),\\
& \alpha_{10} =\frac{1}{2}\rho_2 + \frac{1}{2}\left( \varepsilon_1
-\varepsilon_3 + 2 \varepsilon_4
-  \Tilde{\varepsilon}_1 +  \Tilde{\varepsilon}_4 \right),\\
& \alpha_{11} =\frac{1}{2}\left(\rho_1- \rho_2  \right)
+\frac{1}{2}\left(\varepsilon_1 + \varepsilon_4
+ \Tilde{\varepsilon}_3 + \Tilde{\varepsilon}_4 \right),\\
& \alpha_{12} =\frac{1}{2}\left(- \rho_1- 2 \rho_2  \right)
+\frac{1}{2}\left(-\varepsilon_3 + \varepsilon_4 + \Tilde{\varepsilon}_1 +
\Tilde{\varepsilon}_3 \right),
\end{aligned}
\end{equation}
with an intersection matrix
\begin{equation}
I_{\Z_6}=\text{diag}\left( \left(\begin{array}{cc} 0 & 1\\ -1 & 0
\end{array}\right),
 \cdots, \left(\begin{array}{cc} 0 & 1\\ -1 & 0 \end{array}\right)
\right).
\end{equation}
The D6-branes of the IIA orientifold theory which we are going to
consider can wrap fractional 3-cycles. The limitation on these
cycles is determined by the O6-planes which will be discussed in
the following section.


\subsection{Orientifold projection}\label{Subsec:OProjections}

The aim of this work is to find stable and supersymmetric models.
In order to cancel the RR charge of the D6-branes, orientifold
6-planes are required. Such O6-planes arise naturally, if the
worldsheet parity $\Omega$ is chosen to be accompanied by an
antiholomorphic involution $\R$, which we can choose to be the
complex conjugation
\begin{equation}
\R:\qquad z^k \rightarrow \ov{z}^{k}\qquad \text{for } k=1,2,3\ ,
\end{equation}
where $z^k=x^{2+2k}+i x^{3+2k}$ are the complex coordinates on
every 2-torus $T^2_k$.

\subsubsection{Orientifold images of bulk cycles}\label{Subsubsec:Oplanes}

In order to be consistent with the compactification, $\R$ has to
be an automorphism of the $\Z_6$ invariant lattice. This leads to
two possible orientations {\bf A} and {\bf B} of each 2-torus with
a lattice basis and its dual which are given explicitly in
appendix~\ref{AppSec:BasisTori}. In terms of the notation of
figure~\ref{Fig:Z6torifixedpoints}, the different orientations
lead to the following projections of the fundamental 1-cycles
under $\R$,
\begin{equation}\label{Eq:ORprojectionbulkcycles}
{\bf A}:   \left\{\begin{array}{l}
\pi_{2k-1} \stackrel{\R}{\longrightarrow} \pi_{2k-1}, \\
\pi_{2k} \stackrel{\R}{\longrightarrow} \pi_{2k-1} - \pi_{2k},
\end{array}\right. \qquad \qquad
{\bf B}:  \pi_{2k-1} \stackrel{\R}{\longleftrightarrow}\pi_{2k}.
\end{equation}
The geometry of the first two 2-tori is identical because the
orbifold generator acts in the same way. Consequently, only six of
the $2^3$ naive choices are inequivalent while the choices {\bf
AB} and {\bf BA} for $T^2_1 \times T^2_2$ lead to identical
results. Evaluating the
projection~(\ref{Eq:ORprojectionbulkcycles}) on the right hand
side of~(\ref{Eq:bulkcycles}) leads to the orientifold images of
bulk cycles displayed in table~\ref{Tab:RimagesbulkcyclesZ6}.
\renewcommand{\arraystretch}{1.3}
\begin{table}[ht]
  \begin{center}
    \begin{equation*}
      \begin{array}{|c||c|c|} \hline
        \multicolumn{3}{|c|}{\rule[-3mm]{0mm}{8mm}
\text{\bf $\R$ images of bulk cycles}} \\ \hline\hline
\text{lattice}  &  \R:\rho_1 &  \R:\rho_2 \\ \hline\hline {\bf
AAA} &\rho_1 & \rho_1 - \rho_2 \\\hline {\bf AAB} & \rho_2 &
\rho_1\\\hline {\bf ABA} & \rho_2 & \rho_1\\\hline {\bf ABB} &
\rho_2 - \rho_1 & \rho_2 \\\hline {\bf BBA} & \rho_2 - \rho_1 &
\rho_2\\\hline {\bf BBB} & -\rho_1 & \rho_2 - \rho_1 \\\hline
      \end{array}
    \end{equation*}
  \end{center}

\caption{$\R$ images of cycles inherited from the torus in the
orbifold limit
  $T^6/(\Z_6 \times \OR)$.}
\label{Tab:RimagesbulkcyclesZ6}
\end{table}

The bulk cycles which are invariant under $\R$ are easily read off
from table~\ref{Tab:RimagesbulkcyclesZ6}. However, in order to
determine the homology classes of the O6-planes and thereby the
required sets of D6-branes, the factorizable 3-cycles have to be
considered in more detail. The O6-planes can be decomposed into
two orbits which are invariant under $\OR\theta^{2k}$ and
$\OR\theta^{2k+1}$. The corresponding wrapping numbers, the
coefficients $Y$ and $Z$ of the homological cycles (see
eq.~(\ref{Eq:bulkCoeffYZ})), and bulk cycles are listed in
table~\ref{Tab:O6planesZ6}. The over-all cycle which is wrapped by
the orientifold plane is given by the sum of the two orbits.
\renewcommand{\arraystretch}{1.3}
\begin{table}[ht]
  \begin{center}
    \begin{equation*}
      \begin{array}{|c||c||c|c||c|} \hline
        \multicolumn{5}{|c|}{\rule[-3mm]{0mm}{8mm}
\text{\bf O6-planes for } T^6/(\Z_6 \times \OR)} \\ \hline\hline
\text{lattice}  & (n_1,m_1;n_2,m_2;n_3,m_3) & Y & Z & \text{cycle}
\\ \hline\hline
{\bf AAA} & (1,0;1,0;1,0) & 1 & 0 & \rho_1 \\
& (1,1;1,1;1,-1) & 3 & 0 & 3\rho_1 \\\hline
{\bf AAB} & (1,0;1,0;1,1) & 1 & 1 & \rho_1 + \rho_2 \\
& (1,1;1,1;2,-1) & 3 & 3 & 3(\rho_1 + \rho_2) \\\hline
{\bf ABA} & (1,0;1,1;1,0) & 1 & 1 & \rho_1 + \rho_2 \\
& (1,1;0,1;1,-1) & 1 & 1 & \rho_1 + \rho_2 \\\hline
{\bf ABB} & (1,0;1,1;1,1) & 0 & 3 & 3 \rho_2\\
& (1,1;0,1;2,-1) &  0 & 3 & 3 \rho_2\\\hline
{\bf BBA} & (1,1;1,1;1,0) & 0 & 3 & 3 \rho_2\\
& (0,1;0,1;1,-1) & 0 & 1 & \rho_2\\\hline
{\bf BBB} & (1,1;1,1;1,1) & -3 & 6 & 3(-\rho_1+2\rho_2) \\
& (0,1;0,1;2,-1) & -1 & 2 & -\rho_1+2\rho_2 \\\hline
      \end{array}
    \end{equation*}
  \end{center}
\caption{ O6-planes for $T^6/(\Z_6 \times \OR)$. In each case, the
wrapping numbers of an arbitrary element of a $\Z_6$ orbit are
listed. The coefficients $Y, Z$ are computed
from~(\ref{Eq:bulkCoeffYZ}). The total homology class of the
O6-planes is given by the sum over the two orbits for each
lattice.} \label{Tab:O6planesZ6}
\end{table}

\subsubsection{Orientifold images of exceptional cycles}\label{Subsubsec:OProjex}

In order to find the correct transformations of the fractional
cycles and thereby the D6-branes under the orientifold projection,
the transformations of the relevant fixed points on $T^2_1 \times
T^2_2$ under $\R$ have to be taken into account,
\begin{equation}\label{Eq:Rprojectionfixedpoints}
    {\bf A}:   \left\{\begin{array}{l}
    1 \stackrel{\R}{\longrightarrow} 1,\\
    4  \stackrel{\R}{\longrightarrow} 4,\\
    5 \stackrel{\R}{\longleftrightarrow} 6,
    \end{array}\right. \qquad \qquad
    {\bf B}:  \left\{\begin{array}{l}
    1 \stackrel{\R}{\longrightarrow} 1,\\
    4 \stackrel{\R}{\longleftrightarrow} 5,\\
    6  \stackrel{\R}{\longrightarrow} 6,
    \end{array}\right.
\end{equation}
and the 1-cycle on $T^2_3$ transforms according
to~(\ref{Eq:ORprojectionbulkcycles}).

The images under the reflection $\R$ for all inequivalent lattices
are displayed in table~\ref{Tab:RimagesexccyclesZ6}.
\renewcommand{\arraystretch}{1.15}
\begin{table}[ht]
\small
  \begin{center}
    \begin{equation*}
      \begin{array}{|c||c|c|c|c|c||c|c|c|c|c|} \hline
        \multicolumn{11}{|c|}{\rule[-3mm]{0mm}{8mm}
\text{\bf $\R$ images of exceptional cycles for } T^6/(\Z_6 \times
\OR)} \\ \hline\hline \text{lattice}  &  \R:\varepsilon_1 &
\R:\varepsilon_2 & \R:\varepsilon_3 &  \R:\varepsilon_4 &
\R:\varepsilon_5 &  \R:\Tilde{\varepsilon}_1 &
\R:\Tilde{\varepsilon}_2 & \R:\Tilde{\varepsilon}_3 &
\R:\Tilde{\varepsilon}_4 &  \R:\Tilde{\varepsilon}_5\\
\hline\hline {\bf AAA} & \varepsilon_1 &  \varepsilon_2 &
\varepsilon_3 & \varepsilon_5 &  \varepsilon_4 & \varepsilon_1
-\Tilde{\varepsilon}_1&  \varepsilon_2-\Tilde{\varepsilon}_2 &
\varepsilon_3-\Tilde{\varepsilon}_3 &
\varepsilon_5-\Tilde{\varepsilon}_5 &
\varepsilon_4-\Tilde{\varepsilon}_4 \\\hline {\bf AAB} &
\Tilde{\varepsilon}_1 &  \Tilde{\varepsilon}_2 &
\Tilde{\varepsilon}_3 & \Tilde{\varepsilon}_5 &
\Tilde{\varepsilon}_4 &\varepsilon_1 &  \varepsilon_2 &
\varepsilon_3 & \varepsilon_5 &  \varepsilon_4 \\\hline {\bf ABA}
& \varepsilon_1 & \Tilde{\varepsilon}_2-\varepsilon_2 &
\varepsilon_4 & \varepsilon_3 &
 \varepsilon_5
& \varepsilon_1 -\Tilde{\varepsilon}_1 & \Tilde{\varepsilon}_2 &
\varepsilon_4-\Tilde{\varepsilon}_4 &
\varepsilon_3-\Tilde{\varepsilon}_3 &
\varepsilon_5-\Tilde{\varepsilon}_5\\\hline {\bf ABB} &
\Tilde{\varepsilon}_1 & -\varepsilon_2 & \Tilde{\varepsilon}_4 &
\Tilde{\varepsilon}_3 & \Tilde{\varepsilon}_5 & \varepsilon_1 &
\Tilde{\varepsilon}_2-\varepsilon_2 & \varepsilon_4 &
\varepsilon_3 & \varepsilon_5\\\hline {\bf BBA} &
\Tilde{\varepsilon}_1 -\varepsilon_1 & \Tilde{\varepsilon}_2
-\varepsilon_2 & \Tilde{\varepsilon}_3 -\varepsilon_3 &
\Tilde{\varepsilon}_5 -\varepsilon_5 & \Tilde{\varepsilon}_4
-\varepsilon_4
& \Tilde{\varepsilon}_1 & \Tilde{\varepsilon}_2 &
\Tilde{\varepsilon}_3 & \Tilde{\varepsilon}_5 &
\Tilde{\varepsilon}_4
\\\hline
{\bf BBB} &  -\varepsilon_1 & -\varepsilon_2 &  -\varepsilon_3 &
-\varepsilon_5& \ -\varepsilon_4
& \Tilde{\varepsilon}_1 -\varepsilon_1 & \Tilde{\varepsilon}_2
-\varepsilon_2 & \Tilde{\varepsilon}_3 -\varepsilon_3 &
\Tilde{\varepsilon}_5 -\varepsilon_5 & \Tilde{\varepsilon}_4
-\varepsilon_4
\\\hline
      \end{array}
    \end{equation*}
  \end{center}
\caption{$\R$ images of the exceptional cycles for $T^6/(\Z_6
\times \OR)$.} \label{Tab:RimagesexccyclesZ6}
\end{table}

\subsubsection{RR tadpole cancellation}\label{Subsubsec:RRtadpoles}

The most important consistency requirement on the cohomology
classes of the forms which live on the world volume of the
D6-branes is that the charge of the RR 7-form which couples to the
D6-branes and O6-planes vanishes. The Poincar\'e duals of these
7-forms are homology classes (see e.g.~\cite{Aldazabal:2000dg}),
implying that the RR tadpole cancellation conditions can be
reformulated in terms of the bulk and exceptional 3-cycles
discussed above.

The O6-planes only wrap bulk 3-cycles whereas D6-branes can
wrap both exceptional and bulk 3-cycles. The general condition of
an over-all vanishing homology class is given by
\begin{equation} \label{Eq:Tadp_general}
\sum_a N_a \left( \Pi_a + \Pi_{a'} \right) - 4 \Pi_{O6}=0,
\end{equation}
and can be evaluated in detail using
tables~\ref{Tab:RimagesbulkcyclesZ6}
and~\ref{Tab:RimagesexccyclesZ6} where by $\Pi_{a'}$ we denote the
$\R$ image of the 3-cycle $\Pi_a$. For the six inequivalent
lattice orientations and only bulk branes we obtain
\begin{equation}\label{Eq:RRtadpoles}
\begin{aligned}
{\bf AAA: } &
\sum_a N_a \left( 2 Y_a + Z_a \right) \rho_1 = 2^4 \rho_1,\\
{\bf AAB: } &
\sum_a N_a  \left( Y_a + Z_a \right) \left(\rho_1+\rho_2 \right)= 2^4\left(\rho_1+\rho_2 \right),\\
{\bf ABA: } &
\sum_a N_a  \left( Y_a + Z_a \right) \left(\rho_1+\rho_2 \right)=  2^3\left(\rho_1+\rho_2 \right),\\
{\bf ABB: } &
\sum_a N_a  \left( Y_a + 2 Z_a \right)\rho_2 = 3 \cdot 2^3\rho_2, \\
{\bf BBA: } &
\sum_a N_a  \left( Y_a + 2 Z_a \right)\rho_2 = 2^4\rho_2, \\
{\bf BBB: } & \sum_a N_a Z_a \left(-\rho_1+2 \rho_2 \right) = 2^4
\left(-\rho_1+2 \rho_2 \right).
\end{aligned}
\end{equation}
Here, $N_a$ is the number of D6-branes wrapping the cycle $\Pi_a$,
the so-called stacksize. These conditions are easily generalized
for fractional D6-branes with a bulk part $\frac{1}{2}(Y_a \rho_1
+Z_a \rho_2)$ by inserting the corresponding factor $1/2$ on the
left hand side of~(\ref{Eq:RRtadpoles}).

For exceptional cycles, the over-all homology class has to cancel
among all D6-branes and their $\R$-images by themselves, because
the O6-plane does not contribute.

The same results can be obtained by computing the loop channel
annulus, M\"obius strip and Klein bottle amplitude and performing
the modular transformation to the RR-tree channel. For more
details on the open string amplitudes and massless spectrum see
section~\ref{Sec:spectrum}.


\subsection{Supersymmetry conditions}\label{Subsec:SUSYcond}

\subsubsection{Supersymmetric bulk cycles}\label{Subsubsec:SUSYbulk}

Factorizable 3-cycles preserve ${\cal N}=1$ supersymmetry provided that
the sum over the three angles w.r.t. the $\R$ invariant plane on
all 2-tori vanishes. It is convenient to reformulate the
supersymmetry condition in terms of the coefficients $Y$, $Z$ (see
eq.~(\ref{Eq:bulkCoeffYZ})) as follows. If $\pi \varphi^k_a$ is
the angle w.r.t. $\pi_{2k-1}$ of the bulk cycle $\Pi_a$ represented
by the wrapping numbers $(n^a_k,m^a_k)$, we have
\begin{equation}\label{Eq:wrappingtangent}
\tan(\pi \varphi^k_a) = \sqrt{3}\frac{m_k^a}{2n_k^a+m_k^a}.
\end{equation}
On an {\bf A} lattice, $\pi \varphi^k_a$ is also the angle w.r.t.
the $\R$ invariant axis, whereas on a {\bf B} lattice it is given
by $\pi \varphi^k_a-\frac{\pi}{6}$. We can combine both
possibilities in the single equation $\pi \tilde{\varphi}^{k}_a=
\pi(\varphi^k_a-\frac{b_k}{6})$, where $b_k=0$ for an {\bf A} and
$b_k=1$ for a {\bf B} torus. The necessary condition for a
supersymmetric factorizable bulk cycle is given in terms of the
tangents by $\sum_{k=1}^3 \tan(\pi \Tilde{\varphi}^{k}_a)=
\prod_{k=1}^3 \tan(\pi \Tilde{\varphi}^{k}_a)$. Evaluating this
equation for the six inequivalent choices of lattice orientations
and using~(\ref{Eq:wrappingtangent}), we obtain
\begin{equation}\label{Eq:SUSY1}
\begin{aligned}
{\bf AAA: } & Z_a =0, \\
{\bf AAB: } & Y_a -Z_a=0,\\
{\bf ABA: } & Y_a -Z_a=0,\\
{\bf ABB: } & Y_a =0, \\
{\bf BBA: } & Y_a =0, \\
{\bf BBB: } & 2 Y_a + Z_a =0.
\end{aligned}
\end{equation}
By the analysis of the angle criterion, we recover the result
obtained by stating that the supersymmetric cycles are those
special Lagrangian cycles which are calibrated w.r.t. the same
holomorphic 3-form as the ones wrapped by the O6-planes. Stated
differently, the vanishing coefficients in~(\ref{Eq:SUSY1}) are
also those which are zero in table~\ref{Tab:O6planesZ6}.
Therefore, up to normalization the supersymmetric D6-branes wrap
the same bulk 3-cycle as the O6-planes, and by this the
intersection between the two always vanishes,
\begin{equation}
\Pi_a \circ \Pi_{O6}=0 .
\end{equation}

There remains a little subtlety: since the tangent is periodic in
$\pi$, the condition above does not distinguish between D6-branes
and anti-D6-branes. However, a D6-brane gives a positive
contribution to the untwisted RR charge on the left hand side
of~(\ref{Eq:RRtadpoles}). The sufficient second condition for a
supersymmetric bulk 3-cycle therefore reads
\begin{equation}\label{Eq:SUSY2}
{\bf AAA, AAB, ABA: } Y_a > 0, \qquad  {\bf ABB, BBA, BBB: } Z_a >
0.
\end{equation}


\subsubsection{Supersymmetry condition on exceptional cycles}\label{Subsubsec:SUSYex}

Fractional branes preserve supersymmetry provided that their
contribution from the exceptional cycles arises only from fixed
points on the first two 2-tori which are traversed by the
supersymmetric bulk part of the 2-cycle times a 1-cycle on
$T^2_3$. All possible combinations of factorizable 2-cycles on
$T^2_1 \times T^2_2$ and $\Z_2$ fixed points which they traverse
are displayed in table~\ref{AppTab:WrapFPZ6}. The corresponding
exceptional cycles can be read off from
table~\ref{AppTab:FPandCycles}.

In all cases where the bulk part of the cycle passes through the
origin of $T^2_1 \times T^2_2$, the signs of the contributions
from the three non-trivial fixed points are arbitrary. However, if
the bulk cycle does not pass through the origin, exceptional
cycles arise from four non-trivial fixed points. In this
case, three signs can be chosen independently while the fourth one
is determined to be the product of the other three. This is due to
the fact that relative Wilson lines $\frac{\tau_k}{2} \in
\{0,1/2\}$ between  two branes are associated to the
$\Z_2$ fixed points on $T^2_k$. Since in our convention, the fixed
points are localized on $T^2_1 \times T^2_2$, discrete Wilson
lines naturally occur on these two tori (compare also with~\cite{Cvetic:2000st}). We
choose the convention such that  for vanishing Wilson lines all
fixed points contribute with the same sign. The relative signs
between exceptional 2-cycles are then given as follows,
\begin{equation}\label{Eq:FixpointsWilsonlines}
\tau_0 \left( e_{ik} + (-1)^{\tau_1} e_{jk} + (-1)^{\tau_2}
e_{il} +
  (-1)^{\tau_1+\tau_2}  e_{jl}\right)
=\tau_0 e_{ik} +  \tau^{\prime}_1 e_{jk} + \tau^{\prime}_2  e_{il}
+ \tau_0 \tau^{\prime}_1 \tau^{\prime}_2  e_{jl},
\end{equation}
where $\tau_0 = \pm 1$ is the global sign of contributions from
exceptional cycles corresponding to the two possible $\Z_2$
eigenvalues and $\tau^{\prime}_k=\tau_0(-1)^{\tau_k}$ for $k=1,2$.
The allowed relative signs for combinations of exceptional
3-cycles associated to a specific bulk cycle are obtained by means
of table~\ref{AppTab:FPandCycles}.


\section{Massless open string spectrum: tree and loop amplitudes}\label{Sec:spectrum}
The chiral spectrum and RR tadpole contributions can be computed
from the 3-cycles the D6-branes wrap. However, to ensure that this
method reproduces the string theory calculation it is crucial to
understand the correspondence between the 1-loop open string
amplitudes which allow also for the computation of the non-chiral
spectrum. The cycles are in direct correspondence to the boundary
and crosscap states of the tree channel amplitude, and the 1-loop
expressions are obtained via worldsheet duality.

A stack of $N_a$ identical fractional D6-branes in general
supports the gauge group $U(N_a)$ and preserves ${\cal N}=2$
supersymmetry by itself. This is in contrast to the
compactifications on $T^6$~\cite{Blumenhagen:2000wh,
Blumenhagen:2000ea, Ibanez:2001nd},
$T^6/\Z_3$~\cite{Blumenhagen:2001te}, $T^6/(\Z_3 \times \Z_3)$,
$T^6/(\Z_2 \times \Z_2)$~\cite{Cvetic:2001tj} and $T^6/(\Z_4
\times \Z_2)$~\cite{Honecker:2003vq} which admit only pure bulk
3-cycles and hence preserve ${\cal N}=4$ supersymmetry in the
gauge sector. However, also on $T^6/\Z_6$, additional chiral
multiplets in the adjoint representation arise at intersections of
branes with their orbifold images as explained below.

The chiral part of the open string spectrum can be directly
derived from the intersections of the 3-cycles in a given
configuration as displayed in
table~\ref{Tab:GenericChiralSpectrum},
\renewcommand{\arraystretch}{1.3}
\begin{table}[ht]
  \begin{center}
    \begin{equation*}
      \begin{array}{|c|c|} \hline
\text{multiplicity} & \text{rep.}\\\hline
\Pi_a \circ \Pi_b &  ({\bf N}_a,\ov{\bf N}_b) \\
\Pi_a \circ \Pi_{b'} &   ({\bf N}_a,{\bf N}_b)\\
\frac{1}{2}\left(\Pi_a \circ \Pi_{a'}-\Pi_a \circ \Pi_{O6} \right) &  {\bf Sym}_a \\
\frac{1}{2}\left(\Pi_a \circ \Pi_{a'}+\Pi_a \circ \Pi_{O6} \right)
&  {\bf Anti}_a\\\hline
      \end{array}
    \end{equation*}
  \end{center}
\caption{Generic chiral spectrum of intersecting D6-branes.}
\label{Tab:GenericChiralSpectrum}
\end{table}
whereas the knowledge of the non-chiral states such as  Higgs
particles in a standard model compactification or multiplets in
the adjoint representation requires a detailed analysis of the
string amplitudes or the computation of the Chan-Paton label for
each massless open string state. Both techniques are briefly
discussed in this section in order to show manifestly that our
choice of Chan-Paton matrices and the sign of the orientifold
projection on the exceptional cycles are appropriate. Furthermore,
by using both techniques simultaneously, ambiguities in relative
signs can be eliminated.


\subsection{Boundary states for fractional intersecting branes}\label{Subsec:Boundarystates}

The untwisted and twisted boundary states are directly related to
the bulk and exceptional cycles wrapped by fractional D6-branes.
After applying a modular transformation on the scattering
amplitudes between two D6-branes and between an O6-plane and a
D6-brane, one obtains the two open string 1-loop amplitudes, the
annulus and the M\"obius strip. From these two, we can read off
the complete massless spectrum.

The boundary state of an arbitrary fractional D6-brane consists of
an untwisted part $|U_a\rangle$, corresponding to the wrapped bulk
cycle $\Pi_a^{bulk}$, and a twisted part $|T_a\rangle$, coming
from the exceptional cycles $\Pi_a^{ex}$ stuck at the traversed
$\Z_2$ fixed points,
\begin{equation}
|B_a\rangle =|U_a\rangle +|T_a\rangle.
\end{equation}
For a pure bulk D6-brane, the complete boundary state is given by
\begin{equation}\label{Eq:UntwistedBounState}
|U_a\rangle =c_U \left(\prod_{k=1}^3 L^a_k\right) \left(
\sum_{l=0}^2 |D6;\theta^l (n^a_i,m^a_i)\rangle\right),
\end{equation}
where in $c_U$  all universal factors of the normalization have
been absorbed.

\noindent $L^a_k=\sqrt{(n_k^a)^2+n_k^a m_k^a + (m_k^a)^2}$ is the
length of the 1-cycle wrapping $T^2_k$ measured in units of its
radius $R_k$, and $\theta^l (n^a_i,m^a_i)$ labels a factorizable
3-cycle with wrapping numbers $(n^a_i,m^a_i)$ $(i=1,2,3)$ and its
orbifold images.

A fractional brane wraps only $\frac{1}{2}\Pi_a^{bulk}$.
Accordingly,
 the normalization constant of the untwisted boundary state changes,
$c_U \rightarrow c_U/2$, and the twisted parts of the boundary
states are of the form
\begin{equation}\label{Eq:TwistedBounState}
|T_a\rangle =c_T L^a_3 \sum_{i,j}\alpha_{ij}\left(
\sum_{l=0}^2|D6;\theta^l
 (n^a_3,m^a_3),\theta^l(e_{ij})\rangle\right)
\end{equation}
where all universal factors of the normalization have been
absorbed in $c_T$. The relative factor $c_T/c_U$ is fixed by
worldsheet duality. The twisted boundary states depend only on the
wrapping numbers on $T^2_3$ and are stuck at those $\Z_2$ fixed
points $e_{ij}$  $(i,j=1,4,5,6)$ on $T^2_1 \times T^2_2$ which are
traversed by the bulk cycle. The relative signs $\alpha_{ij}=\pm
1$ for different fixed points correspond to the $\Z_2$ eigenvalue
and discrete  Wilson lines displayed in
equation~(\ref{Eq:FixpointsWilsonlines}).

The oscillator expansion of the boundary states and zero mode
contributions in the annulus amplitude including discrete Wilson
lines are stated in appendix~\ref{AppSec:Amplitude}.

The untwisted crosscap states are constructed in a similar way to
the boundary states. Since the total homology class of the
O6-planes is composed of two independent orbits as shown in
table~\ref{Tab:O6planesZ6}, the crosscap state contains two kinds
of contributions,
\begin{equation}\label{Eq:crosscapcont}
|C\rangle = {\cal N}_{even}\left(\sum_{k=0}^2|\OR\theta^{2k}
\rangle\right) + {\cal N}_{odd}\left(\sum_{k=0}^2|\OR\theta^{2k+1}
\rangle\right),
\end{equation}
where the normalizations ${\cal N}_{even}$, ${\cal N}_{odd}$
depend on the choice of the lattice orientation and can be deduced
from worldsheet duality using the explicit calculation
in~\cite{Blumenhagen:1999ev}. In particular, for an {\bf AB}
lattice on $T^2_1 \times T^2_2$ they are identical.

From~(\ref{Eq:UntwistedBounState}),~(\ref{Eq:TwistedBounState})
and~(\ref{Eq:crosscapcont}), the  tree channel annulus
$\int_0^{\infty} \langle B_a|e^{-2\pi H_{c}l}|B_b \rangle$ and
M\"obius strip $\int_0^{\infty} (\langle B_a|e^{-2\pi H_{c}l}|C\rangle + h.c.$
amplitudes can be computed and then
transformed to the loop channel.


\subsection{Loop channel amplitudes}\label{Subsec:LoopChannel}

The open string 1-loop amplitudes are given by
\begin{equation}\label{Eq:AnnulusMoebius}
\au + \msu = c \int_0^{\infty} \frac{dt}{t^3} \text{Tr}_{open}
\left({\bf P}_{orb}{\bf P}_{GSO}{\bf P}_{\OR}(-1)^{\bf S} e^{-2\pi
t L_0}\right)
\end{equation}
with the projectors ${\bf P}_{orb}=\frac{1}{6} \sum_{k=0}^{5}
\theta^k$, ${\bf P}_{GSO}=\frac{1}{2}(1+(-1)^F)$ and ${\bf
P}_{\OR}=\frac{1}{2}(1+\OR)$ and $\text{Tr}_{open}$ running over
the NS and R sector weighted by $(-1)^{\bf S}$ where ${\bf S}$ is
the space-time fermion number. The massless spectrum can be read
off from~(\ref{Eq:AnnulusMoebius}) by a power series expansion.

In the following, we restrict to supersymmetric configurations and
compute only the number of fermionic degrees of freedom.


\subsubsection{Adjoint and bifundamental representations: Annulus}\label{Subsubsec:Annulus}

A general element of the orbifold group exchanges D6-branes with
their images. Only a $\Z_2$ twist preserves the brane
configuration, such that the only non-vanishing contributions to
the annulus amplitude for bulk branes are given by
\begin{equation}
\au =\frac{c}{24}\int_0^{\infty} \frac{dt}{t^3} \text{Tr}_{open}
\left((1+\theta^3)(1+(-1)^F)(-1)^{\bf S} e^{-2\pi t L_0}\right).
\end{equation}
The generic form of lattice contributions to the amplitudes per
2-torus is displayed in~(\ref{Eq:looplatticecontr}) and  the
oscillator contributions are listed in~(\ref{Eq:OscLoopAM}).

Since the R sector with $(-1)^F$ insertion always has a vanishing
contribution to the loop channel amplitudes, the massless
fermionic spectrum for bulk D6-branes at generic angles
$\pi \varphi^{ab}_k$ is computed from
\begin{equation}\label{Eq:AnnulusR}
\au_R =\sum_{a,b} \frac{c}{24}\int_0^{\infty} \frac{dt}{t^3}
\left((2N_a)(2N_b)I_{ab}
\au^{1/2,0}_{0,(\varphi_1^{ab},\varphi_2^{ab},\varphi_3^{ab})} +
\text{tr}\gamma^a_{\theta^3}\text{tr}\gamma^{b,-1}_{\theta^3}
I^{\theta^3}_{ab}
\au^{1/2,0}_{v=1/2,(\varphi_1^{ab},\varphi_2^{ab},\varphi_3^{ab})}\right),
\end{equation}
where $I_{ab}=\prod_{k=1}^3(n^a_k m^b_k - n^b_k m^a_k)$ is the
intersection number between the factorizable D6-branes $a$ and $b$
and $I^{\theta^3}_{ab}$ is the number of those intersections which
are invariant under the $\theta^3$ insertion, i.e. intersections
localized at the $\Z_2$ fixed points $e_{ij}$ on $T^2_1 \times
T^2_2$ and with arbitrary position on $T^2_3$. One subtlety arises
from the existence of the discrete Wilson lines: the $\Z_2$
invariant intersections are counted with relative signs which come
from the Wilson lines. The matrices $\gamma_{\theta^3}$ are
displayed below in section~\ref{Subsubsec:ChanPaton},
eq.~(\ref{Eq:GammaMatrices}). For bulk branes,
$\text{tr}\gamma^a_{\theta^3}=N_a-N_a=0$, and $\Z_2$ insertions
give vanishing contributions to the annulus amplitude. For
fractional branes $a$, the coefficient of the contribution from
the $\unity$ insertion decreases by a factor of two, i.e. replace
$(2N_a) \rightarrow N_a$, in accord with the expectation from the
boundary state approach
 and $\text{tr}\gamma^a_{\theta^3}$ is replaced by $\tau^a_0 N_a$ where $\tau^a_0=\pm 1$
distinguishes the $\Z_2$ eigenvalues of the two fractional branes
$\frac{1}{2}(\Pi^{bulk}_a \pm \Pi^{ex}_a)$ forming a bulk brane.

For branes parallel on a 2-torus, in the above formula, the
intersection number on the corresponding 2-torus is replaced by
the appropriate lattice sum~(\ref{Eq:looplatticecontr}) and the
oscillator expression is modified as explained in
appendix~\ref{AppSubsec:OsciCont}.

The annulus amplitude is sufficient to compute the massless
spectrum of strings which are not invariant under the orientifold
projection. This comprises the adjoint representations localized
on a stack of identical branes or stuck at intersections of
orbifold images as well as bifundamental representations at
intersections of different branes $a$ and $b$.

For $N_a$ factional branes $a$ which are not their own $\OR$
image, the R sector annulus contributions containing adjoint
representations are given by
\begin{equation}
\begin{aligned}
\au^{R}_{aa} = & \frac{c}{24}\int_0^\infty \frac{dt}{t^3} \Bigl(
N_a^2 \lu^{\au,a}_1 \lu^{\au,a}_2 \lu^{\au,a}_3
\au^{1/2,0}_{0,(0,0,0)}
+(\tau^a_0)^2 N_a^2 \lu^{\au,a}_3 \au^{1/2,0}_{v=1/2,(0,0,0)}\Bigr)\\
&= \frac{c}{24}\int_0^\infty \frac{dt}{t^3} \left( 16N_a^2 +\mathcal{O} (e^{-2\pi t}) \right),\\
\au^{R}_{a(\Theta a)}= & \frac{c}{24}\int_0^\infty \frac{dt}{t^3}
\Bigl( N_a^2 I_{a(\Theta a)} \au^{1/2,0}_{0,(1/3,1/3,-2/3)}
+(\tau^a_0)^2 N_a^2I^{\theta^3}_{a(\Theta a)}
{\cal A}^{1/2,0}_{v=1/2,(1/3,1/3,-2/3)}\Bigr)\\
& = \frac{c}{24}\int_0^\infty \frac{dt}{t^3} \Bigl(( 2I_{a(\Theta a)}+2I^{\theta^3}_{a(\Theta a)})N_a^2 +\mathcal{O} (e^{-2\pi t/3})\Bigr), \\
\au^{R}_{a(\Theta^2 a)}= &  \frac{c}{24}\int_0^\infty
\frac{dt}{t^3} \Bigl( N_a^2 I_{a(\Theta^2 a)}
\au^{1/2,0}_{0,(-1/3,-1/3,2/3)} +(\tau^a_0)^2
N_a^2I^{\theta^3}_{a(\Theta^2 a)}
{\cal A}^{1/2,0}_{v=1/2,(-1/3,-1/3,2/3)}\Bigr)\\
& = \frac{c}{24}\int_0^\infty \frac{dt}{t^3} \Bigl((2
I_{a(\Theta^2 a)}+2I^{\theta^3}_{a(\Theta^2 a)})N_a^2 +\mathcal{O}
(e^{-2\pi t/3}) \Bigr).
\end{aligned}
\end{equation}
It follows that the gauge group with support on $N_a$ identical
branes wrapping $\frac{1}{2}(\Pi^{bulk}_a \pm \Pi^{ex}_a)$ is
$U(N_a)$. The gauge sector preserves ${\cal N}=2$ supersymmetry,
i.e. one multiplet in the adjoint representation is living on the
worldvolume of the stack of branes. This multiplet carries the
degrees of freedom which corresponds to a parallel displacement of
the 1-cycle on the third torus. At each intersection of $a$ with
its orbifold images $(\theta^k a)$, one further chiral multiplet
in the adjoint representation is stuck which is due to the fact
that the orbifold images can recombine into a smooth cycle.


\subsubsection{Symmetric and antisymmetric matter: M\"obius strip}\label{Subsubsec:Moebius}

Arbitrary string configurations are not invariant under $\OR$.
However, some D6-branes can wrap the same 3-cycles as the
O6-planes. In this case, the M\"obius strip gives non-vanishing
contributions to the gauge degrees of freedom and the resulting
gauge group is  special orthogonal.

Furthermore, strings can stretch between a brane $a$ and an
orbifold image in the orbit of the $\R$ image $a'$. These strings
provide for further
 antisymmetric or symmetric representations
of the unitary gauge factor with support on $a$.

The only non-vanishing R sector contributions to the M\"obius
strip are of the form
\begin{equation}
\begin{aligned}
\msu_{R} &= \frac{c}{24}\int_0^\infty \frac{dt}{t^3} \Bigl\{
\mbox{Tr}^{R,aa'}_{\mbox{\scriptsize open}}\left( \OR
(\unity+\theta^3)e^{-2\pi t L_0}\right)
\Bigr. \\
\Bigl. &+\mbox{Tr}^{R,a(\theta a)'}_{\mbox{\scriptsize open}}
\left( \OR (\theta+\theta^4)e^{-2\pi t L_0}\right)
+\mbox{Tr}^{R,a(\theta^2 a)'}_{\mbox{\scriptsize open}} \left( \OR
(\theta^2+\theta^5)e^{-2\pi t L_0}\right) \Bigr\}.
\end{aligned}
\end{equation}
They can be rewritten in terms of oscillator contributions and
eigenvalues $\tau^a_k=\pm 1$ under $\OR\theta^k$ as follows,
\begin{equation}
\msu_{R} =\frac{c}{24}\int_0^\infty \frac{dt}{t^3} \sum_{k=0}^2
\sum_{l=0,1} \tau^a_{k+3l} N_a I^{\R\theta^{k+3l}}_{a(\theta^k
a)'}
\msu_{(\Tilde{\varphi}_1+(k+3l)/6,\Tilde{\varphi}_2+(k+3l)/6,\Tilde{\varphi}_3-(k+3l)/3)}
\end{equation}
where $\pi\Tilde{\varphi}_k$ is the angle of brane $a$ w.r.t. the
$\R$ invariant axis as described in
section~\ref{Subsubsec:SUSYbulk}. $I^{\R\theta^{k+3l}}_{a(\theta^k
a)'}$ counts the number of intersections between $a$ and
$(\theta^k a)'$  which are invariant under $\R\theta^{k+3l}$.

For example, let $a$ and $b$ be the fractional branes wrapping the
bulk parts of the cycles on top of the O6-planes in the {\bf AAB}
model with wrapping numbers as displayed in
table~\ref{Tab:O6planesZ6}, i.e.
$\Pi_a=\frac{1}{2}(\rho_1+\rho_2)+\frac{\tau_0^a}{2}
(\varepsilon_1+\Tilde{\varepsilon}_1+\varepsilon_2+\Tilde{\varepsilon}_2+\varepsilon_3+\Tilde{\varepsilon}_3)$
and $\Pi_b=\frac{3}{2}(\rho_1+\rho_2)-\frac{\tau_0^b}{2}
(\varepsilon_1+\Tilde{\varepsilon}_1+\varepsilon_2+\Tilde{\varepsilon}_2+\varepsilon_3+\Tilde{\varepsilon}_3)$
with $\tau^{a,b}_0= \pm 1$ being the two possible $\Z_2$ eigenvalues.
The $aa$ and $bb$ strings support the gauge group $SO(N_a) \times
SO(N_b)$ and one chiral multiplet in the antisymmetric
representation, ${\bf Anti}_a+{\bf Anti}_b$. Furthermore,
intersections of $a$ with its orbifold images $(\theta^k a)$
($k=1,2$) provide for three multiplets  in ${\bf Anti}_a$, and
$b(\theta^k b)$ strings contribute 15 multiplets in ${\bf
Anti}_b$. For $\tau_0^a=\tau_0^b$, $ab$ and $a(\theta^2 b)$
strings carry three multiplets in $({\bf N}_a,{\bf N}_b)$, and
four multiplets in $({\bf N}_a,{\bf N}_b)$ live on the
$a(\theta b)$ strings. For
$\tau_0^a=-\tau_0^b$, all sectors  $a(\theta^k b)$ give vanishing contributions
to the spectrum.
Observe that this spectrum can be compared to the one
arising from the branes $c$ and $e$ in table~\ref{Tab:3gen_nonchiral_spectrumAAB321}.
In the latter case, the discrete relative Wilson on $T^2_1$
leads to a non-vanishing contribution to the spectrum although
the $\Z_2$ eigenvalues differ.


\subsubsection{Computation of the open spectrum from Chan-Paton labels}\label{Subsubsec:ChanPaton}

The open string spectrum at fractional D6-branes can be computed
in a similar way as for bulk branes using the decomposition into
irreducible representations of the orbifold group as
follows.\footnote{See e.g.~\cite{Bertolini:2003iv} for a recent
review on
  such decompositions in orbifold compactifications.}

One choice of matrices consistent with the fact that the D6-branes
$a$ and $b$ at the end of section~\ref{Subsubsec:Moebius} carry
orthogonal gauge factors and consequently $\OR$ has to preserve
the $\Z_2 $ eigenvalues is given by
\begin{equation}\label{Eq:GammaMatrices}
\gamma_{\theta^3} = \left(\begin{array}{cc} \unity_N & 0 \\ 0 &
-\unity_N
\end{array}\right), \qquad
\gamma_{\OR} = \unity_{2N},
\end{equation}
where $N$ is the number of identical bulk branes. This choice
confirms the global sign of the $\OR$ projection on exceptional
cycles in table~\ref{Tab:RimagesexccyclesZ6}, i.e. $\OR$ acts
merely as a permutation on exceptional 2-cycles but does not bring
about any internal reflection of the blow-ups.

For an arbitrary kind of D6-branes wrapping a bulk 3-cycle which
is not $\OR\theta^k$ invariant, the resulting gauge group is
$U(N)_1 \times U(N)_2$. The two gauge factors belong to the two
fractional branes whose superposition gives the bulk brane.

In particular, the $aa$ strings decompose into $\Z_2$ even
massless states $\psi^I_{-1/2}|0\rangle_{NS}$ for $I=\mu,3,\ov{3}$
and their superpartners which provide for the vector multiplet
carrying the gauge group and one chiral multiplet in the adjoint
representation. For $I=1, \ov{1}, 2, \ov{2}$ the massless states
are $\Z_2$ odd and correspond to strings stretching between the
two fractional branes of opposite $\Z_2$ eigenvalues forming a
bulk brane.

In the example at the end of section~\ref{Subsubsec:Moebius}, only
one fractional cycle of each kind $a$ and $b$ occurs and therefore
the $\Z_2$ odd states do not contribute to  the massless spectrum
on the brane. The representations of states localized at the
intersections $a(\theta^k a)$ and $b(\theta^k b)$ with $k=1,2$ are
computed from the fact that in this case the massless NS and R
states are non-degenerate and $\Z_2$ invariant. The bulk parts of
the cycles $a$ and $(\theta^k a)$ intersect in three $\Z_2$
invariant points while $b$ and $(\theta^k b)$ intersect in three
$\Z_2$ invariant points and 24 points which form pairs under $\R$.
Similarly, the $ab$ and $a(\theta^2 b)$ massless states are
non-degenerate and $\Z_2$ invariant while $a(\theta b)$ massless
string states are twofold degenerate. By counting
the respective intersections, the spectrum in
section~\ref{Subsubsec:Moebius} is recovered.

\section{Anomalies and Green-Schwarz mechanism}\label{Sec:Anomalies_GS}
As discussed in the previous section,
the chiral spectrum and RR tadpole cancellation can be computed either via the 3-cycles or the string loop ampltidues,
and the non-chiral spectrum by means of the latter method. The non-Abelian gauge factors remain unbroken in the
low energy field theory, but the $U(1)$ factors can have anomalies and acquire a mass via couplings to closed string
RR fields. In this section, we discuss this generalised Green Schwarz mechansim and show how to compute the
surviving massless $U(1)$ factors.

The closed string sector contains for all six different choices of
lattice orientations the axion as untwisted RR scalar and five
additional RR scalars from the $\theta^3$ twisted sector. The number
of vectors arising from the RR sectors depends on the lattice.
The complete bosonic closed string spectrum is listed in
table~\ref{Tab:ClosedSpectrumZ6}, the fermionic degrees of freedom follow
from supersymmetry.
\renewcommand{\arraystretch}{1.3}
\begin{table}[ht]
\footnotesize
  \begin{center}
    \begin{equation*}
      \begin{array}{|c||c|c||c|c||c|c||c|c||c|c||c|c|} \hline
        \multicolumn{13}{|c|}{\rule[-3mm]{0mm}{8mm}
\text{\bf Closed string spectrum $T^6/(\Z_6 \times \OR)$}} \\
\hline\hline \text{lattice}
&\multicolumn{2}{|c|}{\rule[-3mm]{0mm}{8mm}{\bf AAA}}
&\multicolumn{2}{|c|}{\rule[-3mm]{0mm}{8mm}{\bf AAB}}
&\multicolumn{2}{|c|}{\rule[-3mm]{0mm}{8mm}{\bf ABA}}
&\multicolumn{2}{|c|}{\rule[-3mm]{0mm}{8mm}{\bf ABB}}
&\multicolumn{2}{|c|}{\rule[-3mm]{0mm}{8mm}{\bf BBA}}
&\multicolumn{2}{|c|}{\rule[-3mm]{0mm}{8mm}{\bf BBB}} \\\hline
\text{sector} & NSNS & RR & NSNS & RR & NSNS & RR & NSNS & RR &
NSNS & RR & NSNS & RR\\\hline\hline \text{untwisted}
&\multicolumn{12}{|c|}{\rule[-3mm]{0mm}{8mm} \text{NSNS: Graviton
+ Dilaton + 8 scalars; RR: Axion + 1 vector}}   \\\hline

\theta+\theta^5 & 4 \text{ s.} & 1 \text{ v.} & 6 \text{ s.} & - &
4 \text{ s.} & 1 \text{ v.} & 6 \text{ s.} & - & 4 \text{ s.} & 1
\text{ v.} & 6 \text{ s.} & -  \\\hline
\theta^2+\theta^4 &20 \text{ s.} & 5 \text{ v.} & 30\text{ s.} & -
& 18 \text{ s.} & 6 \text{ v.} & 24 \text{ s.} & 3 \text{ v.} & 20
\text{ s.} & 5 \text{ v.} & 30\text{ s.} & - \\\hline
\theta^3 & \multicolumn{12}{|c|}{\rule[-3mm]{0mm}{8mm} \text{NSNS:
15 scalars; RR: 5 scalars + 1 vector}}\\\hline
      \end{array}
    \end{equation*}
  \end{center}
\caption{Closed string spectrum on $T^6/(\Z_6 \times \OR)$:
Counting of the
  bosonic degrees of freedom. The fermionic degrees of freedom follow from
  supersymmetry.
1 s. corresponds  to a real scalar while 1 v. denotes a massless
vector with its two helicities.} \label{Tab:ClosedSpectrumZ6}
\end{table}

The axion and the five twisted RR scalars are those fields which
participate in the generalized Green-Schwarz mechanism. Using the
tables~\ref{Tab:RimagesbulkcyclesZ6}
and~\ref{Tab:RimagesexccyclesZ6}, the 3-cycles can be reexpressed
in terms of $\R$ even and $\R$ odd linear combinations $\eta_i$
and $\chi_j$ ($i,j=0,\ldots,5$), respectively, with the property
$\eta_i \circ \chi_j =-4 \delta_{ij}$ and all other intersections
being trivial. The detailed form of these linear combinations
depends on the choice of the lattice as listed in
tables~\ref{Tab:Revencycles} and~\ref{Tab:Roddcycles}.
\renewcommand{\arraystretch}{1.3}
\begin{table}[ht]
  \begin{center}
    \begin{equation*}
      \begin{array}{|c||c|c|c|c|c|c|} \hline
        \multicolumn{7}{|c|}{\rule[-3mm]{0mm}{8mm}
\text{\bf ${\cal R}$ even cycles for } T^6/\Z_6} \\ \hline\hline
\text{lattice}  & \eta_0 & \eta_1 & \eta_2 & \eta_3 & \eta_4 &
\eta_5
 \\ \hline\hline
{\bf AAA} &  \rho_1 & \eps1 & \eps2 & \eps3 & \eps4+\eps5 &
-\Teps4-\eps5+\Teps5
\\\hline
{\bf AAB} &  \rho_1+\rho_2 & \eps1+\Teps1 & \eps2+\Teps2 &
\eps3+\Teps3 & -\Teps4-\eps5 & -\eps4 -\Teps5
\\\hline
{\bf ABA} &  \rho_1+\rho_2 & \eps1 & -\Teps2 & \eps5 & \eps3+\eps4
& -\Teps3-\eps4+\Teps4
\\\hline
{\bf ABB} & \rho_2 & \eps1+\Teps1 & \eps2-2\Teps2 & \eps5+\Teps5 &
-\Teps3-\eps4 & -\eps3-\Teps4
\\\hline
{\bf BBA} &  \rho_2

& -\Teps1 & -\Teps2 & -\Teps3 & \eps4-\eps5+\Teps5 &
-\Teps4-\Teps5
\\\hline
{\bf BBB} & -\rho_1+2\rho_2 & \eps1-2\Teps1 & \eps2-2\Teps2 &
\eps3-2\Teps3 & \eps4-\eps5 & -\Teps4+\eps5-\Teps5
\\\hline
      \end{array}
    \end{equation*}
  \end{center}
\caption{${\cal R}$ even cycles for $T^6/(\Z_6 \times \OR)$.}
\label{Tab:Revencycles}
\end{table}
\renewcommand{\arraystretch}{1.3}
\begin{table}[ht]
  \begin{center}
    \begin{equation*}
      \begin{array}{|c||c|c|c|c|c|c|} \hline
        \multicolumn{7}{|c|}{\rule[-3mm]{0mm}{8mm}
\text{\bf ${\cal R}$ odd cycles for } T^6/\Z_6} \\ \hline\hline
\text{lattice} & \chi_0 & \chi_1 & \chi_2 & \chi_3 & \chi_4 &
\chi_5 \\ \hline\hline {\bf AAA}  & -\rho_1+2\rho_2 &
-\eps1+2\Teps1 & -\eps2+2\Teps2 & -\eps3+2\Teps3 &
\Teps4-\eps5+\Teps5 & \eps4 -\eps5
\\\hline
{\bf AAB}  &   -\rho_1+\rho_2 & -\eps1+\Teps1 & -\eps2+\Teps2 &
-\eps3+\Teps3 & \eps4-\Teps5 & -\Teps4 +\eps5
\\\hline
{\bf ABA} &   -\rho_1+\rho_2 & -\eps1+2\Teps1 & 2\eps2-\Teps2 &
-\eps5+2\Teps5 & \Teps3-\eps4+\Teps4 & \eps3-\eps4
\\\hline
{\bf ABB} &  -2 \rho_1+\rho_2 & -\eps1+\Teps1 & \eps2 &
-\eps5+\Teps5 & \eps3-\Teps4 & -\Teps3+\eps4
\\\hline
{\bf BBA}  &  -2 \rho_1+\rho_2 & 2\eps1-\Teps1 & 2\eps2-\Teps2 &
2\eps3-\Teps3 & \Teps4-\Teps5 & \eps4-\Teps4+\eps5
\\\hline
{\bf BBB} & -\rho_1 & \eps1 & \eps2 & \eps3 & -\eps4+\Teps4-\Teps5
& \eps4+\eps5
\\\hline
      \end{array}
    \end{equation*}
  \end{center}
\caption{${\cal R}$ odd cycles for $T^6/(\Z_6 \times \OR)$.}
\label{Tab:Roddcycles}
\end{table}

A general 3-cycle and its $\R$ image can be rewritten in terms of
the $\R$ even and odd cycles as follows,
\begin{equation}\label{Eq:cycles_rs_basis}
\Pi_a = \sum_{i=0}^5 (r^i_a \eta_i + s^i_a \chi_i),\qquad \Pi_{a'}
= \sum_{i=0}^5 (r^i_a \eta_i - s^i_a \chi_i),
\end{equation}
with $r^i_a, s^i_a$ multiples of one quarter. The RR tadpole
cancellation condition~(\ref{Eq:Tadp_general}) can be rephrased in
terms of these coefficients as
\begin{equation}\label{Eq:Tad_rs_basis}
\sum_b 2 N_b \vec{r}_b = 4 \vec{r}_{O6}
\end{equation}
with $\vec{r}_{O6}=(r^0_{O6},0,0,0,0,0)^T$ and $r^0_{O6}$ as read
off from table~\ref{Tab:O6planesZ6}. Using the fact that the
intersection number of two arbitrary 3-cycles is given by $\Pi_a
\circ \Pi_b= 2(-\vec{r}_a \cdot \vec{s}_b +\vec{s}_a \cdot
\vec{r}_b)$, and multiplying~(\ref{Eq:Tad_rs_basis}) by $2
\vec{s}_a$ leads to
\begin{equation}\label{Eq:CubicAnomalies}
\begin{aligned}
0 
 &= 4 N_a \vec{s}_a \cdot \vec{r}_a -8 \vec{s}_a \cdot\vec{r}_{O6} + 4 \sum_{b
 \neq a} N_b \vec{s}_a \cdot \vec{r}_b \\
 &= N_a \Pi_a \circ \Pi_{a'} -4 \Pi_a \circ \Pi_{O6}
+  \sum_{b \neq a} N_b\left( \Pi_a  \circ \Pi_b +  \Pi_a  \circ
\Pi_{b'}\right).
\end{aligned}
\end{equation}
As in the toroidal orientifold models with intersecting
D6-branes~\cite{Ibanez:2001nd}, equation~(\ref{Eq:CubicAnomalies})
shows manifestly that the RR tadpole cancellation conditions imply
the disappearance of all cubic gauge anomalies of the generic
spectrum listed in table~\ref{Tab:GenericChiralSpectrum} and
impose analogous conditions for matter charged under $SU(2)$ and
$U(1)$ factors.

The mixed gauge anomalies are of the form
\begin{equation}\label{Eq:MixedAnomalies}
{\cal A}_{U(1)_a - SU(N_b)^2} = 4 \vec{s}_a \cdot \left(\vec{r}_b
N_a +\delta_{ab}  \left(\sum_c N_c \vec{r}_c - 2 \vec{r}_{O6}
\right)\right) C_2(N_b).
\end{equation}
The inner bracket in~(\ref{Eq:MixedAnomalies}) vanishes upon RR
tadpole cancellation. The remaining part which is also present for
$a \neq b$ is compensated by the generalized Green-Schwarz
couplings as follows. Six linearly independent $\OR$ even RR
scalars and their dual two-forms can be defined as the pull back
of the ten dimensional $\Omega$ even 3-form and the $\Omega$ odd
5-form over the $\R$ even and $\R$ odd 3-cycles, respectively,
\begin{equation}
\Tilde{\phi}_i = (4\pi^2\alpha')^{-3/2} \int_{\eta_i} C_3, \qquad
\Tilde{B}^i_2 = (4\pi^2\alpha')^{-3/2}  \int_{\chi_i}C_5.
\end{equation}
The axion of table~\ref{Tab:ClosedSpectrumZ6} corresponds to
$\Tilde{\phi}_0$ while the remaining scalars $\Tilde{\phi}_1,
\ldots, \Tilde{\phi}_5$ represent the five RR scalars from the
$\theta^3$ twisted sector.

The four dimensional effective couplings to the gauge fields are
determined by the coefficients of the 3-cycles
in~(\ref{Eq:cycles_rs_basis}),
\begin{equation}\label{Eq:GScouplings}
 \sum_{i=0}^5 2r^i_b  \int_{M_4}  \Tilde{\phi_i} \text{tr} F_b \wedge F_b, \qquad
 N_a \sum_{i=0}^5 2s^i_a  \int_{M_4}\Tilde{B}_i \wedge \text{tr} F_a,
\end{equation}
the factor of two stemming from the couplings to branes $a$ as
well as their images $a'$. The couplings in~(\ref{Eq:GScouplings})
obviously match those in~(\ref{Eq:MixedAnomalies}) and thus
provide for the cancellation of mixed gauge anomalies.

In order for a $U(1)$ factor $Q=\sum_a x_a Q_a$  to remain
massless, the couplings to all two-forms have to vanish, i.e. the
coefficients $x_a$ have to fulfill
\begin{equation}
\sum_a x_a N_a \vec{s}_a=0.
\end{equation}
In a generic model, at most six Abelian gauge factors can acquire
a mass. However, in a supersymmetric set-up all couplings to
$\Tilde{B}_0$ vanish and only up to five Abelian factors can
become massive.

The non-anomalous $U(1)$s can also be calculated
directly from the homological cycles by determining the
kernel of the matrix~\cite{Ott:2003yv,MarchesanoBuznego:2003hp}
\begin{equation}
M_{aI}=N_a\left(v_a^I-v_a^{I\prime}\right),
\end{equation}
where $v_a^I$ and $v_a^{I\prime}$ are just the twelve coefficients
of the homological basis and the $\OR$-mirror, respectively;
$a$ runs over all stacks of branes.


\section{Supersymmetric models}\label{Sec:SUSYmodels}
After having specified the RR tadpole and supersymmetry
conditions, it is now possible to search for concrete
configurations of fractional D6-branes fulfilling these
conditions.

In this section, we explore systematically what kind of gauge
groups and chiral spectra can be obtained. To do so, we will
proceed in the usual way for a certain number $r$ of stacks, each
consisting of $N_r$ branes. Our main aim is to find a
phenomenologically appealing supersymmetric 3 generation model,
ideally without any exotic chiral matter as compared to the
standard model.

The guiding model for our considerations will be the
non-supersymmetric standard model on the toroidal orientifold of ñ\cite{Ibanez:2001nd}
which might also allow for a
supersymmetric extension. It has the two main features that
firstly, it only contains exactly the standard model matter as
chiral matter (plus right handed neutrinos) and secondly, only in
bifundamental representations of the gauge groups. This model is
realized on four stacks plus possibly hidden branes (which do not
intersect with the standard model branes). The second different
possibility to realize the standard model from~\cite{Blumenhagen:2001te}
does not have a supersymmetric extension
on the $\Z_6$-orientifold for the following reason: in this model,
the right handed $u$, $c$ and $t$ quarks have been realized in the
antisymmetric representation of the $U(3)$ brane and it was not
allowed to have any symmetric representations of this gauge
factor. In the present case, the intersection of any brane cycle
$\Pi_a$ with the orientifold plane $\Pi_{O6}$ has to vanish for a
supersymmetric model, i.e. $\Pi_a \circ \Pi_{O6}=0$. But according
to table \ref{Tab:GenericChiralSpectrum}, this just means that the
number of symmetric and antisymmetric representations on a certain
brane are always the same.

In the past, the model building approaches often have
been rather non-systematic, a fact that we will try to avoid here. To do
so, a computer program has been set up in order to first calculate
the configurations in terms of wrapping numbers $n$ and $m$ which
fulfil the untwisted RR tadpole conditions and the supersymmetry
conditions at the same time, where we have to keep in mind that
only the untwisted RR tadpole conditions get a contribution from
the orientifold planes.

Having specified all such configurations in
a certain range of wrapping numbers, all possible fractional
cycles are constructed after a reduction to one representant
of the orbit. This of course has to be done according to the
specific fixed points that the particular fractional brane passes
through. The possible cycles can be determined from table
\ref{AppTab:WrapFPZ6}, where we have allowed a displacement of the
branes from the origin by
\begin{equation}
\sum_{i=1}^{4}\sigma_i \pi_i \qquad \text{with}\qquad \sigma_i \in
\left\{ 0,\frac{1}{2}\right\}
\end{equation}
on the first two $T^2$, according to the discussion in the
preceding sections. For every given configuration of wrapping
numbers fulfilling the RR tadpole conditions in the untwisted
part, this procedure allows for a number of $16\cdot 2^3=128$
different cycles according to equation
(\ref{Eq:FixpointsWilsonlines}) and table \ref{AppTab:WrapFPZ6} for
every stack. This enhances the model building possibilities by a
factor of $128^r$ (where $r$ again is the number of stacks) as
compared to the case of constructions without fractional cycles
(like in the $\mathbb{Z}_3$ orientifold of
\cite{Blumenhagen:2001te}). The reader should be reminded that the
fulfillment of the untwisted RR tadpole equation actually does not
depend on the specific choices of wrapping numbers $n$ and $m$ for
every stack of branes, but only on the corresponding coefficients
of the two homological cycles $Y_a$ and $Z_a$ which are defined in
equation~(\ref{Eq:bulkCoeffYZ}). In contrast, the construction
of twisted cycles does depend on the numbers $n$ and $m$, but actually
only on the oddness and evenness of the wrapping numbers on the
first two tori. These two facts allow for a more
effective and evolved computer algorithm to handle the amount of
computation.

To get started, we will look for all generally possible two stack
configurations (which for now shall be denoted $a$ and $b$). We
will not require to fulfil the RR tadpole equation at this point,
therefore we do not have to specify the size $N_r$ for any of
these two stacks. In this way we get the most general intersection
numbers which -- most optimistically -- are possible in any $\Z_6$
orientifold construction (with more than one stack). These
intersection numbers make up a finite set, if we insist on
constructions without anti-D-branes, simply because every D-brane
gives a positive contribution which never should be larger than
the absolute value of the contribution from the orientifold plane.
This gives an upper bound on $Y_a$ and $Z_a$ if one assumes the
smallest possible stacksize of $N_a=1$. The computation shows that
the possibilities completely agree on all possible tori. They are
listed in table~\ref{Tab:possible_intersection_numbers} for the
case that the intersections of the two branes with themselves and
their orientifold mirror brane are vanishing, i.e.
$I_{aa}=I_{aa'}=I_{bb}=I_{bb'}=0$\footnote{In fact, the intersection
number of any brane with itself vanishes automatically in four
dimensions.}. This means that we require that there are no
symmetric and antisymmetric representations in the corresponding
open string sectors.
\begin{table}[t]
\small
\centering
\sloppy
\renewcommand{\arraystretch}{1.3}
\begin{tabular}{|c|c|}
  \hline

  \multicolumn{2}{|c|}{$(I_{ab}\ , I_{ab'})$}\\
 \hline
  \hline
    $(\phantom{\pm}0\ ,\phantom{\pm}0)$
  &
  $(\pm 1\ ,\pm 1)$ \\

  $(\pm 2\ ,\pm 2)$
  &
  $(\pm 3\ ,\pm 3)$ \\

  $(\pm 5\ ,\pm 5)$
  &
  $(\phantom{\pm}0\ ,\pm 1)$ \\

  $(\pm 1\ ,\phantom{\pm}0)$
  &
  $(\phantom{\pm}0\ ,\pm 3)$ \\

  $(\pm 3\ ,\phantom{\pm}0)$
  &
  $(\pm 1\ ,\pm 3)$ \\

  $(\pm 3\ ,\pm 1)$
  & \\
  \hline
\end{tabular}
\caption{All possible intersection numbers $(I_{ab},I_{ab'})$
between two stacks of branes $a$ and $b$ for intersections with only
bifundamental representations of the gauge group.} \label{Tab:possible_intersection_numbers}
\end{table}
Another computation shows that exactly the same intersections are
still possible if we additionally insist on $\mathcal{N}=1$ supersymmetry on both stacks.

We can already see from this table, that an identical construction of
the standard model spectrum to the one in~\cite{Ibanez:2001nd}
also will not be possible: one cannot get a pair of intersection
numbers between two stacks of the form $(I_{ab},I_{ab'})=(1\ , 2\
)$. In the model in~\cite{Ibanez:2001nd}, this possibility
was necessary in order to realize the three generations of
left-handed quarks as two $SU(2)$ doublets and one antidoublet. In
this way, no additional (w.r.t. the standard model) $SU(2)$ lepton
doublets were required in order to cancel the $U(2)$ anomaly in the
effective 4-dimensional gauge sector.

This problem seems to persist in the construction on the $\Z_6$
orientifold, but later we will see how it actually can be overcome
and that we can even profit from this fact. At this point, it just shall be
mentioned that this might only be a problem in a construction with
four observable intersecting (plus hidden) stacks.

Before coming to the most interesting 5 stack configurations, we
will shortly mention some results on 2, 3 and 4 stack
configurations and how they automatically point towards a 5 stack model.

\subsection{2, 3 and 4 stack configurations}\label{Sec:2_3_4stacks}
Even with 2 stacks of various  stacksizes, already non-trivial
models which fulfill the RR tadpole cancellation conditions and carry matter in bifundamental
 representations can be obtained. But in
all such models on all six different tori, the stacksize of both
stacks has to agree. Besides, there also exists matter in the
symmetric and antisymmetric representation of at least one gauge
factor and furthermore, only even intersection numbers are
possible. This means that the number of bifundamental
representations is always even, too. A typical example for
$N_a=N_b=3$ on the {\bf AAA}-torus is given in table
\ref{tab:2stackmodel}.

\begin{table}
\small
\centering
\sloppy
\renewcommand{\arraystretch}{1.2}
\begin{tabular}{|c||c|l|c|}
  \hline
   stack & $(n_I,m_I)$  & homology cycle& chiral spectrum\\
  \hline\hline
    U(3)  & $(-3,1;-3,2;-1,1)$ & $\Pi_1={1\over 2}\left(
         7\rho_2 -\varepsilon_1-\varepsilon_3 +2\tilde\varepsilon_1+\tilde\varepsilon_3+\tilde\varepsilon_5\right)$& $2 \times ( \bar{\3},\3 )$\\

    & &$\Pi'_1={1\over 2}\left(7\rho_2+2\varepsilon_1+\varepsilon_4+\varepsilon_5-\tilde\varepsilon_1-\tilde\varepsilon_4\right)$& $1 \times ( \3_{A},
    1 ),\ 1 \times ( \mathbf{6}_{S},
    1 )$\\
\cline{1-3}
    U(3)  & $(0,-1;0,1;-1,1)$ &$\Pi_2={1\over 2}\left(
         \rho_2 -2\varepsilon_1-\varepsilon_4-\varepsilon_5+\tilde\varepsilon_1+\tilde\varepsilon_4\right)$& $1\times (1, \bar{\3}_{A}),\
         1\times (1, \bar{\mathbf{6}}_{S})$\\

    & &$\Pi'_2={1\over 2}\left(\rho_2+\varepsilon_1+\varepsilon_3-2\tilde\varepsilon_1-\tilde\varepsilon_3-\tilde\varepsilon_5\right)$& \\
\hline
\end{tabular}
\caption{The wrapping numbers and homology cycles of the D6-branes
in a 2-stack model containing chiral matter.}
\label{tab:2stackmodel}
\end{table}

These limitations
can be explained by the simple fact that the twisted
homological cycles plus their $\R$-mirrors have to be exactly
opposite to each other for the two stacks. Therefore, one cannot obtain
any 3-generation model like the SU(5)-GUT model in~\cite{Blumenhagen:2001te}
and one has to allow for at least 3
stacks for a phenomenologically appealing model.

There is a simple possibility how to sort out right from the start
which stacksizes are able to give valid models and which are not.
Indeed, only the first two components of the RR-tadpole equation
(\ref{Eq:Tadp_general}) which do not depend on the construction of
fractional cycles, are already enough to rule out most of the
possibilities, because if there are no solutions for a given
stacksize, the fractional cycles cannot change this. On the other
hand, if there are solutions of the untwisted RR-tadpole
components, it is not ensured that the twisted RR-tadpole
components also admit solutions.

Using 3 stacks, it turns out that there are solutions of the
untwisted components of the RR-tadpole equation for the most
appealing given stacksizes $N_1=3, N_2=2, N_3=1$. After the
construction of fractional cycles, it is observed that all existing
solutions with chiral matter contain antisymmetric (and by this
also symmetric) representations of the gauge group, but it is
possible to have this only on the most `harmless' $N_3=1$-stack.
Requiring this additional constraint, it is found that just three
bifundamental representations between the $U(3)$- and the
$U(2)$-stack are not possible, so one cannot obtain 3 left-handed
quark generations.

If one turns over to 4 stacks, is is observed quickly from the
untwisted components of the RR-tadpole equation that the most
favorable configuration $N_1=3, N_2=2, N_3=1, N_4=1$ does not
produce any solutions. However, the two configurations $N_1=4,
N_2=2, N_3=1, N_4=1$ and $N_1=3, N_2=2, N_3=2, N_4=1$ which still
might be acceptable, provide for solutions. Requiring again no
(anti-)symmetric representations on the first brane, it is observed
that just on the {\bf ABB}-torus, there exist solutions with 3
bifundamental representations between the first and the second
brane, i.e. there exist 3 left-handed quarks. This seems rather nice, but in the
further investigation of these models it turns out that there are
always more than three $U_R$ and $D_R$ quarks in the $(\bar{\3},1)$
representation of $SU(3) \times SU(2)$ localized at intersections
 between the $U(3)$ and any one of the $U(1)$ stacks.

From all this, we have to conclude that there have to be at least
five stacks of branes to obtain 3 quark generations
and it indeed turns out that with this number
of stacks, it is possible to obtain a very appealing class of
phenomenological models which will be introduced in the following
section.

\subsection{5 stack configurations}\label{Sec:5stacks}

For the search of 5 stack models, we cannot proceed exactly in the
same way as for 4 stack configurations, because it would require
much too much computer power to got through all possible
constructions with fractional cycles. Therefore, we will alter our
approach in the following way: we fix the stacksize on the first
three stacks to be $N_a=3$, $N_b=2$ and $N_c=1$. Then we require from
the beginning that firstly, there are no (anti-)symmetric
representations of the U(3) gauge factor, secondly, that the
absolute value of intersection numbers are $|I_{ab}+I_{ab'}|\equiv
3$ and $|I_{ac}+I_{ac'}|\equiv 6$. This just means that we demand
(up to conjugation) three left-handed quark generations in the
bifundamental representation $(\3,\mathbf{2})$ and that the sum of
$U_R$ and $D_R$ quark generations in the representation
$(\bar\3,1)$ has to be six.

For all homological cycles fulfilling this condition, we now
search for a variable stacksize on the 4th and 5th torus if (and
for which $N_d$, $N_e$) the two untwisted components of the
RR-tadpole equations can be fulfilled. For the remaining
possibilities, we construct all possible fractional cycles also on
the 4th and 5th stack and then look for the spectra which are
possible. At this point, there is a rather miraculous observation
we have made: if we furthermore require that there are no
(anti-)symmetric representations of any gauge factor, there
remains exactly \textit{one} class of models with a given chiral
spectrum.\footnote{This statement is valid up to an overall minus
sign of the intersection numbers, an exchange of the so-far
equivalent 4th and 5th stack and an exchange between the general
sectors $xy$ and $xy'$, all leading to the same massless spectrum.
The same result was found in~\cite{Ibanez:2001nd}.} This model has
the two stacksizes $N_d=N_e=1$. The same observation has been made
on all three possible tori which give results at all (namely the
{\bf AAB}, {\bf ABA} and {\bf ABB} tori) and the chiral spectra of
these models all agree. Even more astonishingly, this model
resembles almost exactly the non-supersymmetric model
of~\cite{Ibanez:2001nd} with regard to the chiral spectrum. There
is just one difference: on the 5th stack $e$, there are three
additional bifundamental representations. At this point, we should
make another remark, being that all nonvanishing intersection numbers
have an absolute value 3, this seems to be a very aesthetical
feature of this model and most likely could be understood in more
depth directly from the $\Z_6$ symmetry.

There is another observation which completely agrees on the whole
class of possibilities: the third and fifth homological cycles are
$\OR$-invariant. A detailed calculation shows that they both
indeed are $SO(2)$-stacks. The chiral spectrum of this class of
models is shown in table \ref{Tab:3gen_chiral_AAB321} for an
example on the {\bf AAB} torus.

\renewcommand{\arraystretch}{1.3}
\begin{table}[ht]
  \begin{center}
    \begin{equation*}
      \begin{array}{|c||c|c||c|c|c||c|} \hline
        \multicolumn{7}{|c|}{\rule[-3mm]{0mm}{8mm}
\text{\bf Chiral spectrum of 5 stack models with $N_a=3$, $N_b=2$, $N_c=N_d=N_e=1$ }} \\ \hline\hline
& \text{sector} & SU(3)_a \times SU(2)_b \times SO(2)_c \times SO(2)_e & Q_a &
Q_b & Q_d & \frac{1}{3}Q_a+Q_d \\\hline
Q_L & ab' & 3 \times (\ov{\3},\2;1,1) & -1 & -1 & 0 & -\frac{1}{3} \\
U_R, D_R & ac & 3\times (\3,1;\2,1) & 1 & 0 & 0 & \frac{1}{3} \\
L & bd' & 3 \times (1,\2;1,1) & 0 & 1 & 1 & 1\\
E_R, N_R & cd & 3 \times (1,1;\2,1) & 0 & 0 & -1 & -1 \\
& be &  3 \times (1,\2;1,\2) & 0 & 1 & 0 & 0
 \\\hline
\end{array}
    \end{equation*}
  \end{center}
\caption{Chiral spectrum of the model class with gauge group $SU(3)_a \times SU(2)_b
  \times SO(2)_c \times U(1)_d \times SO(2)_e \times U(1)_a \times U(1)_b$.}
\label{Tab:3gen_chiral_AAB321}
\end{table}

The fact that we have two $SO(2)$- instead of $U(1)$-stacks is not a
problem, because it just means that the two fractional branes
coincide with their $\OR$-mirror branes. In the non-chiral
spectrum there is the adjoint representation of both gauge
factors, being related to the unconstrained distance between the
two branes and their mirrors on $T^2_3$. If one gives a VEV to these fields,
then the branes and their mirrors are distinguishable and the
gauge group is a $U(1)$ instead of the $SO(2)$ which indeed has
the same rank. Then any bifundamental of the type
$(\3,\mathbf{2})$, where the $\3$ comes for instance from a $U(3)$
and the $\mathbf{2}$ from the $SO(2)$, splits up into a $(\3,1)$
and a $(\3,-1)$, where $1$ and $-1$ now are the $U(1)$-charges.
After this transition, the $U_R$- and $D_R$-quarks (and the $E_R$
and $N_R$) are distinguishable as usual by their opposite $U(1)$
charge.

Beside these common properties, the non-chiral spectrum which has
been calculated according to the lines of section
\ref{Sec:spectrum} disagrees for different models on just one and
also between different tori. This is well understandable, because
it does not only depend on the homology, but also on the
geometrical properties of the branes and the choice of the
lattice. In the next section, we will discuss two different
explicit examples of this class of models in more detail.
\newpage
\section{The Supersymmetric Standard model}\label{Sec:SUSY-SM}
In the previous section, we systematically explored how it is
possible to obtain three quark generations in a stable and RR
tadpole free D6-brane configuration. In this section, we explore
in detail how the supersymmetric standard model arises.

\subsection{The model on the {\bf AAB} torus}
The explicit configuration that shall be discussed in this section
is given in table \ref{tab:5stackmodel321AABsetup}, the
homological cycles and intersection numbers are listed in table
\ref{tab:5stackmodel321AABhomology}.
\begin{table}
\centering
\sloppy
\renewcommand{\arraystretch}{1.3}
\begin{tabular}{|c||c|c|c|c|}

  \hline
    & $(n_I,m_I)$ & $(\sigma_1,\sigma_2,\sigma_3,\sigma_4)$ & $(\tau_1,\tau_2)$ & $\Z_2$\\
  \hline\hline
    $N_a=3$
         & $(-2,1;-1,2;-2,1)$& $(0,0,0,0)$ & $(0,0)$ & $-$ \\

\hline
    $N_b=2$
         & $(-1,0;-1,1;-1,2)$& $(0,0,0,0)$ & $(1,0)$ & $+$ \\

\hline
    $N_c=1$
         & $(-2,1;-2,1;-1,2)$& $(0,0,0,0)$ & $(1,0)$ & $+$ \\

\hline
    $N_d=1$
         & $(-1,0;-1,1;-1,2)$& $(0,0,0,0)$ & $(0,0)$ & $-$ \\

\hline
    $N_e=1$
         & $(-1,1;-1,1;-2,1)$&  $(0,0,0,0)$ & $(0,0)$ & $-$ \\
         \hline
\end{tabular}
\caption{The setup of the D6-branes
in the 5 stack model on the {\bf AAB} torus.}
\label{tab:5stackmodel321AABsetup}
\end{table}
\begin{table}
\centering
\sloppy
\renewcommand{\arraystretch}{1.3}
\begin{tabular}{|c||c|c|c|c|ll|}
  \hline
\multicolumn{5}{|c|}{\bf homology cycles}& \multicolumn{2}{|c|}{\bf intersections}\\
\hline\hline
\multicolumn{5}{|l|}{ $\Pi_a={1\over 2}\left(
         3\rho_1 +3\rho_2 -\varepsilon_1 +2\varepsilon_2-\varepsilon_5+2\tilde\varepsilon_1-\tilde\varepsilon_2+2\tilde\varepsilon_5\right)$}
         &$I_{ab}=0$ & $I_{ab'}=-3$ \\
\cline{1-5}
\multicolumn{5}{|l|}{ $\Pi_b={1\over 2}\left(
         \rho_1 +\rho_2 +\varepsilon_1 +2\varepsilon_2+\varepsilon_5-2\tilde\varepsilon_1-\tilde\varepsilon_2-2\tilde\varepsilon_5\right)$}
         &$I_{ac}=3$ & $I_{ac'}=3$ \\
\cline{1-5}
\multicolumn{5}{|l|}{ $\Pi_c={1\over 2}\left(
         3\rho_1 +3\rho_2 +\varepsilon_1 -\varepsilon_2+\varepsilon_3+\tilde\varepsilon_1-\tilde\varepsilon_2+\tilde\varepsilon_3\right)$}
         &$I_{bd}=0$ & $I_{bd'}=3$ \\
\cline{1-5}
\multicolumn{5}{|l|}{ $\Pi_d={1\over 2}\left(
         \rho_1 +\rho_2 +\varepsilon_1 -2\varepsilon_2+\varepsilon_5-2\tilde\varepsilon_1+\tilde\varepsilon_2-2\tilde\varepsilon_5\right)$}
         &$I_{cd}=3$ & $I_{cd'}=-3$ \\
\cline{1-5}
\multicolumn{5}{|l|}{ $\Pi_e={1\over 2}\left(
         \rho_1 +\rho_2 -\varepsilon_1 -\varepsilon_2-\varepsilon_3-\tilde\varepsilon_1-\tilde\varepsilon_2-\tilde\varepsilon_3\right)$}
         &$I_{be}=3$ & $I_{be'}=3$ \\
\hline
\end{tabular}
\caption{The homology cycles and intersection numbers (all other intersection numbers vanishing) of the D6-branes
in the 5 stack model on the {\bf AAB} torus.}
\label{tab:5stackmodel321AABhomology}
\end{table}
After the displacement of the
third and fifth stack on the third 2-torus (as discussed in the preceding
section), the chiral spectrum takes the form as shown in table \ref{Tab:3gen_chiral_AAB321afterU1}.
\renewcommand{\arraystretch}{1.3}
\begin{table}[ht]
  \begin{center}
    \begin{equation*}
      \begin{array}{|c||c|c||c|c|c|c|c|l||l|} \hline
        \multicolumn{10}{|c|}{\rule[-3mm]{0mm}{8mm}
\text{\bf Chiral spectrum of 5 stack model on {AAB} torus}} \\ \hline\hline
& \text{sector} & SU(3)_a \times SU(2)_b & Q_a &
Q_b & Q_c & Q_d & Q_e & Q_{B-L}& Q_Y \\\hline
Q_L & ab' & 3 \times (\ov{\3},\2) & -1 & -1 & 0 & 0 & 0 & \frac{1}{3}& \frac{1}{6} \\
U_R& ac & 3\times (\3,1) & 1 & 0 & -1 & 0 & 0 & -\frac{1}{3}& -\frac{2}{3} \\
D_R & ac' & 3\times (\3,1) & 1 & 0 & 1 & 0 & 0 & -\frac{1}{3}& \frac{1}{3}  \\
L & bd' & 3 \times (1,\2) & 0 & 1 & 0 & 1 & 0 & -1& -\frac{1}{2} \\
E_R & cd & 3 \times (1,1) & 0 & 0 & 1 & -1 & 0 &1 & 1\\
N_R & cd' & 3 \times (1,1) & 0 & 0 & -1 & -1 & 0 & 1&0 \\
& be &  3 \times (1,\2) & 0 & 1 & 0 & 0 & -1 & 0& 0\\
& be' &  3 \times (1,\2) & 0 & 1 & 0 & 0 & 1 & 0 & 0\\
\hline
\end{array}
    \end{equation*}
  \end{center}
\caption{Chiral spectrum of the model class with gauge group $SU(3)_a \times SU(2)_b
  \times U(1)_a \times U(1)_b \times U(1)_c \times U(1)_d \times U(1)_e$.}
\label{Tab:3gen_chiral_AAB321afterU1}
\end{table}
We can immediately calculate which $U(1)$ factors remain massless
after applying the Green-Schwarz mechanism which is discussed in
detail in section \ref{Sec:Anomalies_GS}. The result is that one
obtains three $U(1)$s which are free of triangle anomalies, being
$Q_{B-L}=-\frac{1}{3}Q_a-Q_d$, $Q_c$ and $Q_e$. As in similar constructions, the
first one is the $B-L$ symmetry, $Q_c$ is twice the third
component of the right-handed weak isospin. $Q_e$ is an additional
$U(1)$ symmetry under which none of the standard model particles
transforms, only the two additional fields.

Also the linear combination
\begin{equation}
Q_Y=-\frac{1}{6}Q_a+\frac{1}{2}Q_c-\frac{1}{2}Q_d
\end{equation}
is massless, being the hypercharge $Y$. Every chiral field has the
correct quantum numbers (for both hypercharge and $B-L$ symmetry),
the only mystical one being the two additional kinds of fields at the $be$
and $be'$ intersections. Actually, these two kinds of fields have the right quantum
numbers to be the supersymmetric standard model partners of the
Higgs fields with a vanishing hypercharge, $H$ and $\bar
H$.\footnote{The hypercharge could also be defined with an
additional factor of $-1/2Q_e$, then the two additional kinds of fields
would have the more familiar opposite hypercharge $-1/2$ and
$1/2$.} The only problem in the present construction is the fact
that these bifundamental fields do not stretch from the second
stack to the $c$-brane, but to the $e$-brane, and therefore do not
give rise to the standard Yukawa couplings.

But the problem at first sight can be overcome as well. It is
known that in many cases a gauge breaking where two unitary gauge
groups are broken to the diagonal subgroup is possible, in our
case the breaking $U(1)_c\times U(1)_e\rightarrow U(1)_C$ would be
needed. In the language of D-branes, this requires a brane
recombination mechanism which in some cases can be understood in
the effective field theory just as a Higgs effect. We will explain
in the following why this is exactly the case for the present
model.

If two stacks of D-branes preserve a common $\mathcal{N}=2$
supersymmetry, then a massless hypermultiplet being localized at
the intersection indicates that there is a possible deformation of
the two stacks into just one recombined one. In our case, which is
similar to the one described in \cite{Blumenhagen:2002gw}, two
factorizable branes can only preserve a common $\mathcal{N}=2$
supersymmetry if they are parallel on one of the three tori. To
understand this for our concrete model, we have to take a look at
the non-chiral spectrum. This has been calculated for the concrete
model of table \ref{tab:5stackmodel321AABsetup} as described in
section \ref{Sec:spectrum} and the result is shown in table
\ref{Tab:3gen_nonchiral_spectrumAAB321}.
\renewcommand{\arraystretch}{1.2}
\begin{table}[ht]
  \begin{center}
    \begin{equation*}
      \begin{array}{|c|l|c||c|l|c|} \hline
        \multicolumn{6}{|c|}{\rule[-3mm]{0mm}{8mm}
\text{\bf Non-chiral massless and light open spectrum, AAB on } T^6/(\Z_6
\times \OR)} \\ \hline\hline
\text{sector} & {U(3)_a \times U(2)_b}_{(Q_c, Q_d, Q_e)} &  \sqrt{\alpha'}m
& \text{sector} & {U(3)_a \times U(2)_b}_{(Q_c, Q_d, Q_e)} &  \sqrt{\alpha'}m
\\\hline
aa & 16 \times ({\bf 9},1)_{0,0,0} & 0 & ad' & 6 \times [(\3,1)_{0,1,0}  +h.c.] & 0 \\
bb & 4 \times (1,{\bf 4})_{0,0,0} & 0 &  &  (\3,1)_{0,1,0}  +h.c. & \Sigma_{ad'} \\
cc & 16 \times (1,1)_{0,0,0} & 0 & ae &  6 \times [(\3,1)_{0,0,-1} +h.c.] & 0 \\
dd  & 4 \times (1,1)_{0,0,0} & 0 & & (\3,1)_{0,0,-1} +h.c. & \Sigma_{ae} \\
ee &  4 \times (1,1)_{0,0,0} & 0 & ae' & 6 \times  [(\3,1)_{0,0,1} +h.c.] & 0 \\
aa' & 9 \times [(\3_A,1)_{0,0,0}+h.c.] & 0 &  & (\3,1)_{0,0,1} +h.c. & \Sigma_{ae'} \\
 & 5 \times [(\3_A,1)_{0,0,0}+h.c.] & \Sigma_{aa'} &bc & 6 \times [(1,\2)_{-1,0,0} +h.c.]  & 0 \\
bb' & 3 \times [(1,{\bf 1}_A)_{0,0,0}+ h.c.] &  0 &  & (1,\2)_{-1,0,0} +h.c. & \Sigma_{bc} \\
 & (1,{\bf 1}_A)_{0,0,0}+h.c. &  \Sigma_{bb'} & bc' & 6 \times [(1,\2)_{1,0,0} +h.c.]  & 0 \\
 ab & 2 \times [(\3,\ov{\2})_{0,0,0}+h.c.] & \Sigma_{ab} && (1,\2)_{1,0,0} +h.c. & \Sigma_{bc'} \\
ac &  3\times [(\3,1)_{-1,0,0} +h.c.] & 0 &ce & 2 \times [(1,1)_{1,0,-1} +h.c.] & \Sigma_{ce}\\
 &  4\times [(\3,1)_{-1,0,0} +h.c.]& \Sigma_{ac} &ce' & 2 \times [(1,1)_{1,0,1} +h.c.] & \Sigma_{ce'}\\
ac' & 3\times [(\3,1)_{1,0,0} +h.c.] &  0 &de & 3 \times [(1,1)_{0,1,-1} +h.c] & 0 \\
 & 4\times [(\3,1)_{1,0,0} +h.c.]& \Sigma_{ac'} && (1,1)_{0,1,-1} +h.c & \Sigma_{de} \\
ad & 3 \times [(\3,1)_{0,-1,0} +h.c.] & 0 & de' & 3 \times [(1,1)_{0,1,1} +h.c] & 0 \\
 & 4 \times [(\3,1)_{0,-1,0} +h.c.] & \Sigma_{ad} & & (1,1)_{0,1,1} +h.c & \Sigma_{de'} \\\hline
\end{array}
    \end{equation*}
  \end{center}
\caption{Non-chiral massless and light open spectrum of the
model in table~\ref{tab:5stackmodel321AABsetup} on {\bf AAB}
computed from 3-cycles. If
  the factorizable cycles are parallel on $T^2_3$, the mass of the
states is proportional to the relative distance
$\Sigma_{xy}=|\sigma_{56}^x-\sigma_{56}^y|$ of the D6-branes on
this torus, $\sqrt{\alpha'}m_{xy}=\Sigma_{xy}$. The distance has
to be taken to be in the range $\Sigma_{xy} \in [0,1/2] \times R_3
$ with $R_3$ being the (dimensionless) length scale on $T^2_3$.
$\sigma_{56}^x$ denotes the distance of brane $x$ from the origin
on $T^2_3$.
 The $\3_A$ of $U(3)_a$
transforms as $\3_2$ under the decomposition $U(3)a \rightarrow
SU(3)_a \times U(1)_a$, while the $\ov{\2}$ of $U(2)_b$ decomposes into $\2_{-1}$ of $SU(2)_b \times U(1)_b$.
The physical $U(1)$ charges can be
computed from section \ref{Sec:Anomalies_GS}.
}
\label{Tab:3gen_nonchiral_spectrumAAB321}
\end{table}

\begin{figure}[t]
\begin{center}
\includegraphics[scale=1.0]{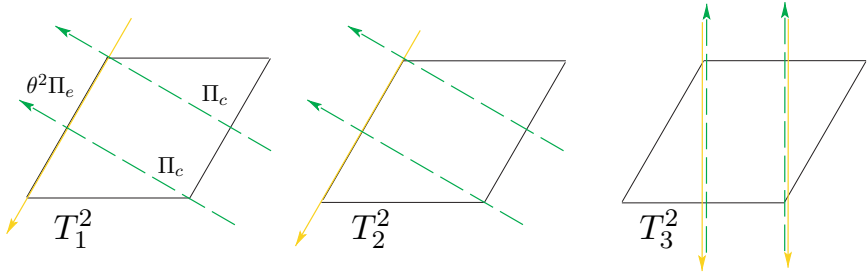}
\end{center}
\caption{Geometrical intersections of branes $c$ and $(\theta^2 e)$
on the {\bf AAB} torus. The gauge group is $U(1)_c \times U(1)_e$ if the
branes are displaced from the origin on $T^2_3$. Remember that the $\R$ invariant plane
lies along $\pi_1 \otimes \pi_3 \otimes (\pi_5+\pi_6)$ with the notation
as in figure~\ref{Fig:Z6torifixedpoints}.} \label{Fig:intersections_aab_c_tqe}
\end{figure}

From this table, we can immediately see that the necessary
hypermultiplets are indeed existing, being the ones in the sector
between the $c$ and $e$-brane. If we take a closer look at the
computation and for a moment distinguish  between the different
orbifold images under the $\theta$ action\footnote{Of course, in
the end we have to sum over all $\theta$ images such that the
result is invariant under the orbifold action.}, we observe that
the two hypermultiplets are  not in the $ce$ sector, but in
the $c\ (\theta^2 e)$ sector. In this sector, the branes $c$ and
$(\theta^2 e)$ are indeed parallel on the third torus, see figure
\ref{Fig:intersections_aab_c_tqe}.
Therefore, we can recombine in this way the $c$ with the
$(\theta^2 e)$-brane (and at the same time the $(\theta c)$ with
the $e$ brane and the $(\theta^2 c)$ with the $(\theta e)$ brane)
by giving a VEV to the two hypermultiplets in the following way:
it is possible to understand the recombination as a Higgs effect
in the $\mathcal{N}=2$ effective theory, if there exists a flat
direction $\langle h_1\rangle=\langle h_2\rangle$ in the D-term
potential
\begin{equation}
V_D\sim\frac{1}{2g^2}\left(h_1 \bar h_1-h_2 \bar h_2\right)^2,
\end{equation}
along which the gauge symmetry is broken to the diagonal subgroup.
Here, $h_1$ and $h_2$ denote the two chiral multiplets inside the
hypermultiplet. If one gauge factor is a $U(1)$, there is a
potential problem: if the two involved stacks intersect only once
(meaning that there is just one hypermultiplet), then the D-flat
direction is not F-flat because there is a superpotential,
coupling the two chiral multiplets to the adjoint vector multiplet
$\Phi$ as
\begin{equation}
W=h_1 h_2 \Phi .
\end{equation}
Therefore $\partial W/\partial \Phi$ imposes $h_1 h_2=0$, meaning
that we cannot give a VEV to both fields $h_1$ and $h_2$ at the
same time (what D-flatness actually requires).

The situation in our case is different: looking at the geometrical
intersections of brane $c$ and $(\theta^2 e)$ in figure
\ref{Fig:intersections_aab_c_tqe}, it can be observed that they
intersect on both the first and the second torus twice, but
hypermultiplets live only on two of the intersections, because the
relative Wilson line on $T^2_1$ projects out the other two.

For two hypermultiplets, there is one flat direction which is
obtained by combining the VEVs for $h_1, h_2$ in one hypermultiplet and
for $\tilde h_1, \tilde h_2$ in the other, such that both the D- and F-flatness
conditions are fulfilled. This means explicitly that
\begin{align}
&W_1=h_1 h_2 \Phi ,\\
&W_2=\tilde h_1 \tilde h_2 \Phi .\nonumber
\end{align}
From F-flatness, it is possible to give $h_2$ and $\tilde h_1$ a
non-vanishing VEV (while at the same time giving a vanishing one to
$h_1$ and $\tilde h_2$) and still obtain a flat direction in the
D-term potential because they couple to the same vectormultiplet
$\Phi$, i.e. the D-term looks like
\begin{equation}
V_D\sim\frac{1}{2g^2}\left(h_1 \bar h_1-h_2 \bar h_2
+\tilde h_1 \bar{\tilde h}_1-\tilde {h}_2 \bar {\tilde h}_2\right)^2\ .
\end{equation}
The D-term and F-terms are indeed flat for the choice
\begin{align}
&\langle h_1\rangle=\langle \tilde{h}_2\rangle=0\ ,\\
&\langle h_2\rangle=\langle \tilde{h}_1\rangle\neq 0\ .
\end{align}
This construction is just possible if the VEVs that we have given
to the fields in between the $c$ and $c'$ brane and the $e$ and
$e'$ brane (formerly the adjoints on the world volume of the $\OR$
invariant stacks, in order to get $U(1)$ instead of $SO(2)$ gauge
groups) are the same, but this is completely unproblematic.

Homologically, we only have to add the two cycles $\Pi_c$ and
$\Pi_e$ to get the recombined brane, which shall be denoted as
$\Pi_C=\Pi_c+\Pi_e$. This complex cycle has the same volume as the
sum of volumes of the two cycles before recombination occurs. The
intersection numbers of the recombined cycles are given in table
\ref{tab:5stack321AABint_recomb}.
\begin{table}
\centering
\sloppy
\renewcommand{\arraystretch}{1.3}
\begin{tabular}{|ll|}
  \hline
\multicolumn{2}{|c|}{\bf intersections}\\
\hline\hline
$I_{ab}=0$ & $I_{ab'}=-3$ \\
$I_{aC}=3$ & $I_{aC'}=3$ \\
$I_{bd}=0$ & $I_{bd'}=3$ \\
$I_{Cd}=3$ & $I_{Cd'}=-3$ \\
$I_{bC}=3$ & $I_{bC'}=3$ \\
\hline
\end{tabular}
\caption{The intersection numbers (all other intersection numbers
vanishing) of the D6-branes in the final 5 stack model on the {\bf AAB}
torus.} \label{tab:5stack321AABint_recomb}
\end{table}
The final chiral spectrum after the recombination is given in
table \ref{Tab:3gen_chiral_AAB321after_rek}.
\renewcommand{\arraystretch}{1.3}
\begin{table}[ht]
  \begin{center}
    \begin{equation*}
      \begin{array}{|c||c|c||c|c|c|c||l|} \hline
        \multicolumn{8}{|c|}{\rule[-3mm]{0mm}{8mm}
\text{\bf Chiral spectrum of 5 stack model on {AAB} torus}} \\ \hline\hline
& \text{sector} & SU(3)_a \times SU(2)_b & Q_a &
Q_b & Q_C & Q_d & Q_Y\\
\hline
Q_L & ab' & 3 \times (\ov{\3},\2) & -1 & -1 & 0 & 0 & \frac{1}{6}\\
U_R& aC & 3\times (\3,1) & 1 & 0 & -1 & 0 & -\frac{2}{3}\\
D_R & aC' & 3\times (\3,1) & 1 & 0 & 1 & 0 & \frac{1}{3}\\
L & bd' & 3 \times (1,\2) & 0 & 1 & 0 & 1 & -\frac{1}{2}\\
E_R & Cd & 3 \times (1,1) & 0 & 0 & 1 & -1 &1\\
N_R & Cd' & 3 \times (1,1) & 0 & 0 & -1 & -1 &0\\
H& bC &  3 \times (1,\2) & 0 & 1 & -1 & 0 & -\frac{1}{2}\\
\bar H& bC' &  3 \times (1,\2) & 0 & 1 & 1 & 0 & \frac{1}{2}\\
\hline
\end{array}
    \end{equation*}
  \end{center}
\caption{Chiral spectrum of the model class with gauge group $SU(3)_a \times SU(2)_b
  \times U(1)_a \times U(1)_b \times U(1)_C \times U(1)_d$.}
\label{Tab:3gen_chiral_AAB321after_rek}
\end{table}
The massless $U(1)$s can be checked again using the computation of
the Green-Schwarz couplings along the lines of
chapter~\ref{Sec:Anomalies_GS}. There are two non-anomalous
$U(1)$s, being $Q_{B-L}=-\frac{1}{3}Q_a-Q_d$ and $Q_C$.
Fortunately, the hypercharge
\begin{equation}
Q_Y=-\frac{1}{6}Q_a+\frac{1}{2}Q_C-\frac{1}{2}Q_d
\end{equation}
is still massless and the two types of fields $H$ and $\bar H$ have just
the opposite hypercharge $-1/2$ and $1/2$. Therefore, they can be
exactly understood as the superpartners of the Standard model Higgs
with a definite chirality.

\subsection{The model on the {\bf ABA} torus}
The explicit configuration that shall be discussed in this section
is given in table \ref{tab:5stackmodel321ABAsetup}, the
homological cycles and intersection numbers are listed in table
\ref{tab:5stackmodel321ABAhomology}.

\begin{table}
\centering
\sloppy
\renewcommand{\arraystretch}{1.3}
\begin{tabular}{|c||c|c|c|c|}

  \hline
    & $(n_I,m_I)$ & $(\sigma_1,\sigma_2,\sigma_3,\sigma_4)$ & $(\tau_1,\tau_2)$ & $\Z_2$\\
  \hline\hline
    $N_a=3$
         & $(-1,1;-1,1;-2,1)$ & $(0,0,0,0)$ & $(0,0)$ & $-$ \\

\hline
    $N_b=2$
         & $(-1,1;-1,1;-2,1)$ & $(0,0,0,0)$ & $(0,1)$ & $-$ \\

\hline
    $N_c=1$
         & $(-1,0;-1,1;-1,2)$ & $(0,\frac{1}{2},0,0)$ & $(1,0)$ & $-$ \\

\hline
    $N_d=1$
         & $(-1,1;-1,1;-2,1)$ &  $(0,0,0,0)$ & $(0,0)$ & $+$ \\

\hline
    $N_e=1$
         & $(-1,0;-1,1;-1,2)$ & $(0,\frac{1}{2},0,0)$ & $(1,1)$ & $-$  \\
         \hline
\end{tabular}
\caption{The setup of the D6-branes
in the 5 stack model on the {\bf ABA} torus.}
\label{tab:5stackmodel321ABAsetup}
\end{table}
\begin{table}
\centering
\sloppy
\renewcommand{\arraystretch}{1.3}
\begin{tabular}{|c||c|c|c|c|ll|}
  \hline
\multicolumn{5}{|c|}{\bf homology cycles}& \multicolumn{2}{|c|}{\bf intersections}\\
\hline\hline
\multicolumn{5}{|l|}{ $\Pi_a={1\over 2}\left(
         \rho_1 +\rho_2 -\varepsilon_1 -\varepsilon_2-\varepsilon_3-\tilde\varepsilon_1-\tilde\varepsilon_2-\tilde\varepsilon_3\right)$}
         &$I_{ab}=0$ & $I_{ab'}=3$ \\
\cline{1-5}
\multicolumn{5}{|l|}{ $\Pi_b={1\over 2}\left(
         \rho_1 +\rho_2 -\varepsilon_1 +\varepsilon_2+\varepsilon_3-\tilde\varepsilon_1+\tilde\varepsilon_2+\tilde\varepsilon_3\right)$}
         &$I_{ac}=-3$ & $I_{ac'}=-3$ \\
\cline{1-5}
\multicolumn{5}{|l|}{ $\Pi_c={1\over 2}\left(
         \rho_1 +\rho_2 +3\varepsilon_1 +2\varepsilon_3+\varepsilon_4-\tilde\varepsilon_3+\tilde\varepsilon_4\right)$}
         &$I_{bd}=0$ & $I_{bd'}=-3$ \\
\cline{1-5}
\multicolumn{5}{|l|}{ $\Pi_d={1\over 2}\left(
         \rho_1 +\rho_2 +\varepsilon_1 +\varepsilon_2+\varepsilon_3+\tilde\varepsilon_1+\tilde\varepsilon_2+\tilde\varepsilon_3\right)$}
         &$I_{cd}=-3$ & $I_{cd'}=3$ \\
\cline{1-5}
\multicolumn{5}{|l|}{ $\Pi_e={1\over 2}\left(
         \rho_1 +\rho_2 +3\varepsilon_1 -2\varepsilon_3-\varepsilon_4+\tilde\varepsilon_3-\tilde\varepsilon_4\right)$}
         &$I_{be}=-3$ & $I_{be'}=-3$ \\
\hline
\end{tabular}
\caption{The homology cycles and intersection numbers (all other intersection numbers vanishing) of the D6-branes
in the 5 stack model on the {\bf ABA} torus.}
\label{tab:5stackmodel321ABAhomology}
\end{table}
Comparing table \ref{tab:5stackmodel321ABAsetup} with the model of
the preceding section, table \ref{tab:5stackmodel321AABsetup},
reveals that the realization on the torus is very different, for
instance in the previous case stacks $b$ and $d$ are parallel while in this model stacks $a$, $b$ and $d$
are parallel as well as $c$ and $e$. Furthermore, the stacks $c$ and $e$ are displaced from the
origin on $T^2_1$.

Nevertheless, the chiral spectrum agrees up to a conjugation with
the chiral spectrum of the model on the {\bf AAB} torus, see table
\ref{Tab:3gen_chiral_AAB321afterU1}. The massless $U(1)$s after
the application of the Green-Schwarz formalism do also completely
agree. On the other hand, calculating the non-chiral spectrum for
this second explicit model shows that it is indeed very different
as compared to the one on the {\bf AAB} torus. The spectrum is
listed in table \ref{Tab:3gen_nonchiral_ABA_321}.
\renewcommand{\arraystretch}{1.3}
\begin{table}[ht]
  \begin{center}
    \begin{equation*}
      \begin{array}{|c|c|c||c|c|c|} \hline
        \multicolumn{6}{|c|}{\rule[-3mm]{0mm}{8mm}
\text{\bf Non-chiral massless and light open spectrum, ABA on } T^6/(\Z_6
\times \OR)} \\ \hline\hline
\text{sector} & {U(3)_a \times U(2)_b}_{(Q_c, Q_d, Q_e)}& \sqrt{\alpha'}m
& \text{sector} & {U(3)_a \times U(2)_b}_{(Q_c, Q_d, Q_e)} & \sqrt{\alpha'}m
\\\hline
aa & 4 \times ({\bf 9},1)_{0,0,0} & 0 & ab & 3 \times [(\3,\ov{\2})_{0,0,0}+h.c.] & 0 \\
bb & 4 \times (1,{\bf 4})_{0,0,0} & 0 &ab' & (\3,\2)_{0,0,0}+h.c. & \Sigma_{ab'}\\
cc & 4 \times (1,1)_{0,0,0} & 0 &ac & (\3,1)_{-1,0,0} +h.c. & \Sigma_{ac} \\
dd  & 4 \times (1,1)_{0,0,0} & 0 &ac' & (\3,1)_{1,0,0} +h.c. & \Sigma_{ac'}\\
ee &  4 \times (1,1)_{0,0,0} & 0 &ad & 2 \times [(\3,1)_{0,-1,0} +h.c.] & \Sigma_{ad} \\
aa' & (2 \times \3_A+{\bf 6}_S,1)_{0,0,0}+h.c. & 0  &ad' & 3 \times [(\3,1)_{0,1,0}  +h.c.] & 0 \\
 & (\3_A,1)_{0,0,0}+h.c. & \Sigma_{aa'} & ae &  (\3,1)_{0,0,-1} +h.c. & \Sigma_{ae} \\
bb' & (1,2 \times {\bf 1}_A+\3_S)_{0,0,0}+ h.c. &  0 &ae' &  (\3,1)_{0,0,1} +h.c. & \Sigma_{ae'} \\
 & (1,{\bf 1}_A)_{0,0,0}+h.c. &  \Sigma_{bb'} &bc & (1,\2)_{-1,0,0} +h.c. & \Sigma_{bc} \\
cc' & (1,1)_{2,0,0}+h.c. & 0 &bc' & (1,\2)_{1,0,0} +h.c. & \Sigma_{bc'} \\
dd' & (1,1)_{0,2,0}+h.c. & 0 &be & (1,\2)_{0,0,-1} +h.c. & \Sigma_{be}\\
ee' & (1,1)_{0,0,2}+h.c. & 0 & be' & (1,\2)_{0,0,1} +h.c. & \Sigma_{be'}\\\hline
\end{array}
    \end{equation*}
  \end{center}
\caption{Non-chiral massless and light open spectrum of the model from table~\ref{tab:5stackmodel321ABAsetup} on the {\bf
    ABA} torus. The notation agrees with table
\ref{Tab:3gen_nonchiral_spectrumAAB321}.} \label{Tab:3gen_nonchiral_ABA_321}
\end{table}
Remarkably, the brane recombination mechanism which has been
described in the last section does not work for this model: the
hypermultiplets in the sector between the $c$ and the $e$ brane
are absent due to the relative Wilson line on $T^2_2$.

Therefore, we can draw the conclusion that it is only possible for
a subclass of models with the same initial chiral spectrum to break it
to the exact supersymmetric standard model.
\newpage
\section{Left-right symmetric models}\label{Sec:LR-sym-SUSY-models}

In order to obtain a left-right symmetric model with three quark generations, also at least five stacks
are required. For the lattices {\bf AAA}, {\bf BBA} and {\bf BBB}, it turns out that all brane
configurations with $SU(3)_a \times SU(2)_b \times SU(2)_c$, no (anti)symmetric
chiral states of $SU(3)_a$, i.e. $\Pi_a \circ \Pi_{a'}=0$, and three left and right handed quark generations,
i.e.  $3 \times (\3_a,\2_b)+3\times(\ov{\3}_a,\2_c)$ in the chiral spectrum, wrap
larger bulk cycles than the O6-planes. Therefore, no supersymmetric chiral left-right symmetric
3-generation model on these tori fulfills RR tadpole cancellation.
This result agrees completely with the observation made for the ansatz $N_a=3$, $N_b=2$ and $N_c=1$ in the
previous section.
For the {\bf ABA} and {\bf ABB} lattices, the minimal requirement on three stacks of fractional D6-branes
can be fulfilled. However, the twisted RR charges cannot be cancelled by any configuration with at most
five stacks of D6-branes - at least if we require that the two additional gauge groups have at most rank
two.

For the lattice {\bf AAB}, there exist chiral 3-generation left-right symmetric models with five stacks.
Two distinct chiral spectra occur, one of them containing (anti)symmetric representations of some
$SU(2)$ factors. The other chiral spectrum
encloses only bifundamental representations as displayed in table~\ref{Tab:3gen_chiral_AAB_LR}.
The gauge group of the standard model part consists of
$SU(3)_a \times SU(2)_b \times SU(2)_c \times U(1)_d$. In order to fulfill tadpole cancellation,
an additional stack with gauge group $U(2)_e$ is required.
Apart from the left/right symmetric MSSM particles, two kinds of exotic chiral particles
charged under the additional $U(2)_e$ arise. These
seemingly unwanted exotic particles, however, have the correct quantum
numbers to combine into composite Higgs particles. I.e., also in the
left/right symmetric models the Higgs fields can originate from the chiral spectrum
due to an `internal' $U(2)_e$ symmetry.

\renewcommand{\arraystretch}{1.3}
\begin{table}[ht]
\small
  \begin{center}
    \begin{equation*}
      \begin{array}{|c||c|c||c|c|c|c|c||c|c|} \hline
        \multicolumn{10}{|c|}{\rule[-3mm]{0mm}{8mm}
\text{\bf Chiral left-right symmetric spectrum of an AAB model on } T^6/(\Z_6 \times \OR)} \\ \hline\hline
& \text{sector} & SU(3)_a \times SU(2)_b \times SU(2)_c \times SU(2)_e & Q_a &
Q_b & Q_c& Q_d& Q_e  & Q_{B-L} & \Tilde{Q}  \\\hline
Q_L & ab' & 3 \times (\3,\2,1;1) & 1 & 1 & 0 & 0 & 0 & 1/3 & 2\\
U_R, D_R & ac' & 3\times (\ov{\3},1,\2;1) & -1 & 0 & -1 & 0 & 0 & -1/3 & -2 \\
L & bd' & 3 \times (1,\2,1;1) & 0 & -1 & 0 & -1 & 0 & -1 & -1\\
E_R, N_R & cd' & 3 \times (1,1,\2;1) & 0 & 0 & 1 & 1 & 0 & 1 & 1 \\
 & be' &  3 \times (1,\2,1;\2) & 0 & -1 & 0 & 0 & -1 & 0 & -5/2\\
& ce' & 3 \times (1,1,\2;\2) & 0 & 0 & 1 & 0 & 1 & 0 & 5/2
 \\\hline
\end{array}
    \end{equation*}
  \end{center}
\caption{Chiral spectrum of a left-right symmetric model.
In the last two columns, the charges under the two massless linear combinations for the
specific D6-brane configuration~(\ref{Tab:App_AAB_322_conf}) are displayed.}
\label{Tab:3gen_chiral_AAB_LR}
\end{table}

A concrete realisation of this chiral spectrum is given in table \ref{Tab:App_AAB_322_conf}.
\begin{table}[ht]
\centering
\sloppy
\renewcommand{\arraystretch}{1.3}
\begin{tabular}{|c||c|c|c|c|}

  \hline
    & $(n_I,m_I)$ & $(\sigma_1,\sigma_2,\sigma_3,\sigma_4)$ & $(\tau_1,\tau_2)$ & $\Z_2(=\tau_0)$\\

  \hline\hline
    $N_a=3$
         & $(-2,1;-1,2;-2,1)$& $(0,0,0,0)$ & $(0,0)$ & $-$ \\

\hline
    $N_b=2$
         & $(-1,0;-1,1;-1,2)$& $(0,0,0,0)$ & $(0,1)$ & $+$ \\

\hline
    $N_c=2$
         & $(-1,0;-1,1;-1,2)$& $(0,0,0,0)$ & $(1,0)$ & $+$ \\

\hline
    $N_d=1$
         & $(-1,0;-1,1;-1,2)$& $(0,0,0,0)$ & $(0,0)$ & $-$ \\

\hline
    $N_e=2$
         & $(-1,0;-1,1;-1,2)$&  $(0,0,0,0)$ & $(1,1)$ & $+$ \\
         \hline
\end{tabular}
\caption{The setup of the D6-branes
in the 5 stack left/right symmetric model on the {\bf AAB} torus.}
\label{Tab:App_AAB_322_conf}
\end{table}
The corresponding 3-cycles can be read off from tables~\ref{AppTab:WrapFPZ6} and~\ref{AppTab:FPandCycles}.
Two Abelian gauge factors remain massless,
\begin{equation}
Q_{B-L}=\frac{1}{3}Q_a+Q_d, \qquad
\Tilde{Q} = \frac{1}{4} \left(3Q_a+5Q_b+5Q_c-Q_d+5Q_e\right).
\end{equation}
The non-chiral massless spectrum of the configuration~\ref{Tab:App_AAB_322_conf} is listed in
table~\ref{Tab:3gen_compl_spectrum_AAB_322}.
\renewcommand{\arraystretch}{1.3}
\begin{table}[ht]
\small
  \begin{center}
    \begin{equation*}
      \begin{array}{|c|c||c|c|} \hline
        \multicolumn{4}{|c|}{\rule[-3mm]{0mm}{8mm}
\text{\bf Non-chiral left-right massless and light open spectrum, AAB, } T^6/(\Z_6 \times \OR)} \\ \hline\hline
\text{sector} & U(3)_a \times U(2)_b \times U(2)_c \times U(2)_e (\times U(1)_d)
& \text{sector} & U(3)_a \times U(2)_b \times U(2)_c \times U(2)_e (\times U(1)_d)
\\\hline
aa & 16 \times ({\bf 9},1,1;1)_0  & ad & 7 \times [(\3,1,1;1)_{-1} +h.c.] \\
bb & 4 \times (1,{\bf 4},1;1)_0  &ad' & 7 \times [(\3,1,1;1)_{1} +h.c.] \\
cc & 4 \times (1,1,{\bf 4};1)_0   &ae  & 2 \times [(\3,1,1;\ov{\2})_0 +h.c.] \\
dd & 4 \times (1,1,1;1)_0   &ae' & 3 \times [(\3,1,1;\2)_0 +h.c.] \\
ee & 4 \times (1,1,1;{\bf 4})_0 &  bc &  3 \times [(1,\2,\ov{\2};1)_0+h.c.] \\
aa' & 14\times [({\bf 3}_A,1,1;1)_0  +h.c.] & bc' &  3 \times [(1,\2,\2;1)_0+h.c.] \\
bb'& 4\times [(1,{\bf 1}_A,1;1)_0+h.c.] &be &  3 \times [(1,\2,1;\ov{\2})_0+h.c.] \\
cc' &4\times [(1,1,{\bf 1}_A;1  )_0+h.c.] &be' &  (1,\2,1;\2)_0+h.c. \\
ee'& 4\times [(1,1,1;{\bf 1}_A  )_0+h.c.] &ce &  3 \times [(1,1,\2;\ov{\2})_0+h.c.] \\
ab & 2 \times [(\3,\ov{\2},1;1)_0 +h.c.] &ce' &  (1,1,\2;\2)_0+h.c. \\
ac & 2 \times [(\3,1,\ov{\2};1)_0 +h.c.] & & \\\hline
\end{array}
    \end{equation*}
  \end{center}
\caption{Non-chiral massless and light open spectrum of the left/right symmetric model computed from cycles.}
\label{Tab:3gen_compl_spectrum_AAB_322}
\end{table}



\section{Conclusions and prospects}\label{Sec:Conclusions}

In this article, we have worked out all technical details of
computing chiral and non-chiral massless spectra for intersecting
fractional D6-branes on the $\Z_6$ orientifold. Discrete Wilson
lines and distances of branes naturally occur due to the existence
of exceptional cycles. The $\Z_3$ subsymmetry on each two torus
freezes all complex structure moduli. Supersymmetry projects onto
one out of two possible toroidal cycles. In addition, the
D6-branes can wrap some of the ten existing exceptional 3-cycles.
This leads to a ${\cal N}=2$ supersymmetric gauge sector as
compared to the ${\cal N}=4$ ones for toroidal and orbifold
backgrounds without exceptional 3-cycles. In a supersymmetric
set-up, all contributions to the bulk cycles are proportional to
those of the O6-planes. In the $\Z_6$ orientifold, therefore,
non-trivial intersections arise purely from the exceptional part
of fractional branes. This approach also could be generalized to
the two further symmetric orbifold groups $\Z_6 \times \Z_3$ and
$\Z_{6'}$. The former is briefly mentioned in
appendix~\ref{AppSec:Z6Z3}, but does not seem to be of any
phenomenological interest due to the small number of fractional
cycles.

Already for 2-stack configurations, non-trivial chiral spectra
with bifundamental representations exist. By systematically
examining the possible 2, 3 and 4-stack configurations, we find
that they cannot provide for non-trivial chiral spectra with the
Standard model gauge group. In all models, RR tadpoles are
cancelled and supersymmetry is preserved globally, ensuring also
the absence of NS-NS tadpoles.

The first configurations with the Standard model gauge group and
also the correct chiral matter exist for 5 stacks. The most
miraculous fact is that with only making the requirements of
having at least one 3-, one 2- and one 1-stack of branes, having 3
quark generations, no chiral matter in antisymmetric
representations, and preserving globally ${\cal N}=1$ supersymmetry,
there remains only \textit{one} model with a definite chiral
spectrum, which is shown in table~\ref{Tab:3gen_chiral_AAB321afterU1}.

Looking more closely at this model, it is not just very
aesthetical in having only non-vanishing intersection numbers of
an absolute value of 3, beside a massless hypercharge, it also
seems to give rise to exactly the standard model particles in the
chiral spectrum in addition to two additional kinds of particles
in the bifundamental representation $(1,\2)$ of $U(1)\times U(2)$.
This looks like the two superpartners of the Higgs in a
supersymmetric standard model, the only seemingly problem being
the fact that the $U(1)$ is not from the right stack, so at first
sight they cannot give rise to the standard Yukawa couplings.

The non-chiral spectrum is rather different for two representants
of the discussed class of models. However, through a well
motivated brane recombination process which is shown to exactly
correspond to a Higgs branch in the effective ${\cal N}=2$ theory of the
type $U(1)\times U(1)\rightarrow U(1)$, these particles indeed can
be identified with the two kinds of Higgs multiplets in one
concrete realization. In homology, this process just means that we
add the two factorizable 3-cycles for the two involved
$U(1)$-branes to get the recombined one, which then is
non-factorizable. In the second presented example, the
recombination still works in homology, but cannot be understood as
a Higgs effect in the effective field theory, because the
necessary fields in between the two branes are missing.

We have to emphasize again that the model presented in
section~\ref{Sec:SUSY-SM} therefore represents the first
compactification with intersecting D6-branes at angles and
genuinely three generations, i.e. no brane recombination is
required to obtain three quark and lepton families.

A similar reasoning applies to the left-right symmetric model
displayed in section~\ref{Sec:LR-sym-SUSY-models}.

It will be interesting, to explore the whole class of discussed
models with the given chiral spectrum in more detail, meaning that
one could compare all different concrete realizations in a spirit
like in \cite{Douglas:2003um}. Besides, the behavior of the Yukawa
and gauge coupling constants depends on the internal geometry and
the full massless spectrum. For instance, one could try to
calculate the threshold corrections to these models as in
\cite{Lust:2003ky} and determine if gauge coupling unification is
possible \cite{Blumenhagen:2003jy}. Furthermore, supersymmetry
breaking sources should arise at some point and might be
understandable in these models. All these tasks hopefully will be
achieved in the future~\cite{HO:2004}.

\section{Erratum}\label{Sec:Conclusions}
A comparison with the CFT computations
in~\cite{Blumenhagen:1999ev,Blumenhagen:2004di} reveals that the
overall sign of the $\OR$ projection on the exceptional cycles in
table 3 has to be the opposite of the stated one and at the same
time the overall sign of the self-intersection matrix of the
exceptional cycles, equation (15) has to be changed.\footnote{We
thank R. Blumenhagen and J. Conlon for discussions on this point.}

This exchanges the $\OR$ even and odd exceptional cycles $\eta_i$
and $\chi_i$ for $i=1 \ldots 5$ in table 6 and 7. Furthermore, the
overall toroidal cycle wrapped by the D6-branes in the concrete
models presented in section 5 and 6 have to be smaller by a factor
of two, leading to at most rank 8 for the {\bf AAA}, {\bf AAB},
{\bf BBA} and {\bf BBB} orientations and 12 for {\bf ABB} and
globally supersymmetric configurations. Models with higher rank in
the observable sector require the presence of hidden sector
supersymmetry breaking (bulk)-anti-D6-branes. The supersymmetric
chiral spectrum in table 13 is still obtained with the change of
signs and overall length of the O6-planes e.g. on the {\bf AAB}
torus in the set-up
\begin{equation}
\begin{array}{|c||c|c|c|c|}\hline
 & (n_I,m_I) & (\sigma_1, \sigma_2, \sigma_3, \sigma_4) & (\tau_1,\tau_2) & \Z_2 \\\hline\hline
N_a=3 & (-1,0; -1,1; -1,2) &  (0,0,0,0) &  (0,0) & -\\
N_b=2 &  (-1,0; -1,1; -1,2) &   (0,0,0,0) & (1,0) & - \\
N_c=1 &  (-1,1; -1,1; -2,1) &  (0,0,1/2,0) & (0,1) & -\\
N_d=1 &  (-1,0; -1,1; -1,2) &  (0,0,0,0) & (0,0) & +  \\
N_e=1 &  (-1,1; -1,1; -2,1) & (0,0,1/2,0) & (1,1) & - \\\hline
\end{array}
\end{equation}
with the homological cycles
\begin{equation}
\begin{aligned}
 \pi_a &= \frac{1}{2}\left(\rho_1+\rho_2 + \eps1-2\Teps1-2\eps2+\Teps2+\eps5-2\Teps5 \right),\\
 \pi_b &= \frac{1}{2}\left(\rho_1+\rho_2-\eps1+2\Teps1-2\eps2+\Teps2-\eps5+2\Teps5\right),\\
\pi_c &= \frac{1}{2}\left(\rho_1+\rho_2+3\eps2-3\Teps2-\eps4-\Teps4+\eps5+\Teps5\right),\\
 \pi_d &= \frac{1}{2}\left(\rho_1+\rho_2-\eps1+2\Teps1+2\eps2-\Teps2-\eps5+2\Teps5\right),\\
 \pi_e &= \frac{1}{2}\left(\rho_1+\rho_2+3\eps2-3\Teps2+\eps4+\Teps4-\eps5-\Teps5\right).
\end{aligned}
\end{equation}
The cycles $\pi_c$ and $\pi_e$ are again $\OR$ invariant due to
the interplay of the discrete Wilson line and displacement on
$T^2_2$. Also the hypercharge for the model on the {\bf AAB} torus
in section 6 is still massless. On all other tori such a model
cannot be obtained any longer. The conclusions of the paper are
unchanged. \vskip 1cm

\noindent {\bf Acknowledgments}

\noindent It is a pleasure to thank L.~G\"orlich, K.~Landsteiner,
A.~ Uranga, K.~Wendland, J.~Cascales, W.~Troost, A.~van~ Proeyen,
D.~L\"ust and especially R.~Blumenhagen
and L.~Huiszoon for helpful discussions.\\
This work is supported by the RTN programs under contract numbers
HPRN-CT-2000-00131 and HPRN-CT-2000-00148 and in part also by the
Federal Office for Scientific, Technical and Cultural Affairs
through the "Interuniversity Attraction Poles Programme -- Belgian
Science Policy" P5/27.


\begin{appendix}

\section{Basis for a $\Z_3$ invariant 2-torus}\label{AppSec:BasisTori}

We fix the angle between the two basis vectors of a $\Z_3$
invariant 2-torus to be $\pi/3$. With this restriction, the two
possible $\R$ invariant lattices are spanned by
\begin{equation}
\begin{aligned}
e^{\bf A}_1 & = \left(\begin{array}{c} \sqrt{2} \\ 0 \end{array}\right), \qquad
e^{\bf A}_2  = \left(\begin{array}{c} 1/\sqrt{2}\\ \sqrt{3/2} \end{array}\right), \\
e^{\ast{\bf A}}_1 & = \left(\begin{array}{c} 1/\sqrt{2} \\ -1/\sqrt{6} \end{array}\right), \qquad
e^{\ast{\bf A}}_2  = \left(\begin{array}{c} 0 \\ \sqrt{2/3} \end{array}\right),
\end{aligned}
\end{equation}
for the {\bf A} orientation where $\pi_{2k-1}$ lies on the $\R$ invariant plane,
and
\begin{equation}
\begin{aligned}
e^{\bf B}_1 & = \left(\begin{array}{c} \sqrt{3/2}  \\ -1/\sqrt{2} \end{array}\right), \qquad
e^{\bf B}_2  = \left(\begin{array}{c} \sqrt{3/2}  \\  1/\sqrt{2}\end{array}\right), \\
e^{\ast{\bf B}}_1 & = \left(\begin{array}{c} 1/\sqrt{6} \\ -1/\sqrt{2}  \end{array}\right), \qquad
e^{\ast{\bf B}}_2  = \left(\begin{array}{c} 1/\sqrt{6} \\  1/\sqrt{2} \end{array}\right),
\end{aligned}
\end{equation}
for the {\bf B} orientation where $\R$ exchanges $\pi_{2k-1}$ and $\pi_{2k}$.


\section{Exceptional cycles,  wrapping numbers and fixed points
on $T^2_1 \times T^2_2$}\label{AppSec:WrappingsFixedpoints}

In this section, some technical details regarding the fractional
branes are given. Table \ref{AppTab:WrapFPZ6} lists the
corresponding fixed points which are possible to be traversed for
a given set of geometric brane wrapping numbers and displacements.
The connection between the orbits of the traversed fixed points
and the corresponding exceptional cycles is subsequently given in table
\ref{AppTab:FPandCycles}.
\begin{table}[h!]\label{tab:wrapfp}
\scriptsize
  \begin{center}
    \begin{equation*}
      \begin{array}{|c||c|c|c||c|c|c|} \hline
        \multicolumn{7}{|c|}{\rule[-3mm]{0mm}{8mm}
\text{\bf Wrapping numbers intersecting fixed points for } T^6/\Z_6} \\ \hline\hline
(n_2,m_2) & (\odd,\even) & (\even, \odd) & (\odd,\odd)
 & (\odd,\even) & (\even, \odd) & (\odd,\odd)\\ \hline
(n_1,m_1) & \multicolumn{3}{|c|}{\sigma_3=\sigma_4=0}
& \multicolumn{3}{|c|}{\sigma_3=\frac{1}{2}, \sigma_4=0}\\\hline\hline
\multicolumn{7}{|c|}{\sigma_1=\sigma_2=0}\\\hline
(\odd ,\even) &
(e_{11})\; e_{14}\; e_{41}\; e_{44}
&
(e_{11})\; e_{15}\; e_{41}\; e_{45}
&
(e_{11})\; e_{16}\; e_{41}\; e_{46}
&
(e_{11})\; e_{14}\; e_{41}\; e_{44}
&
e_{14}\; e_{16}\; e_{44}\; e_{46}
&
e_{14}\; e_{15}\; e_{44}\; e_{45}
\\\hline
(\even, \odd) &
(e_{11})\; e_{14}\; e_{51}\; e_{54}
&
(e_{11})\; e_{15}\; e_{51}\; e_{55}
&
(e_{11})\; e_{16}\; e_{51}\; e_{56}
&
(e_{11})\; e_{14}\; e_{51}\; e_{54}
&
e_{14}\; e_{16}\; e_{54}\; e_{56}
&
e_{14}\; e_{15}\; e_{54}\; e_{55}
\\\hline
 (\odd,\odd) &
(e_{11})\; e_{14}\; e_{61}\; e_{64}
&
(e_{11})\; e_{15}\; e_{61}\; e_{65}
&
(e_{11})\; e_{16}\; e_{61}\; e_{66}
&
(e_{11})\; e_{14}\; e_{61}\; e_{64}
&
e_{14}\; e_{16}\; e_{64}\; e_{66}
&
e_{14}\; e_{15}\; e_{64}\; e_{65}
\\\hline\hline
\multicolumn{7}{|c|}{\sigma_1=\frac{1}{2}, \sigma_2=0} \\\hline
(\odd,\even) &
(e_{11})\; e_{14}\; e_{41}\; e_{44}
&
(e_{11})\; e_{15}\; e_{41}\; e_{45}
&
(e_{11})\; e_{16}\; e_{41}\; e_{46}
&
(e_{11})\; e_{14}\; e_{41}\; e_{44}
&
e_{14}\; e_{16}\; e_{44}\; e_{46}
&
e_{14}\; e_{15}\; e_{44}\; e_{45}
\\\hline
(\even, \odd) &
e_{61}\; e_{64}\; e_{41}\; e_{44}
&
e_{61}\; e_{65}\; e_{41}\; e_{45}
&
e_{61}\; e_{66}\; e_{41}\; e_{46}
&
e_{61}\; e_{64}\; e_{41}\; e_{44}
&
e_{64}\; e_{66}\; e_{44}\; e_{46}
&
e_{64}\; e_{65}\; e_{44}\; e_{45}
\\\hline
 (\odd,\odd) &
e_{51}\; e_{54}\; e_{41}\; e_{44}
&
e_{51}\; e_{55}\; e_{41}\; e_{45}
&
e_{51}\; e_{56}\; e_{41}\; e_{46}
&
e_{51}\; e_{54}\; e_{41}\; e_{44}
&
e_{54}\; e_{56}\; e_{44}\; e_{46}
&
e_{54}\; e_{55}\; e_{44}\; e_{45}
\\\hline\hline
\multicolumn{7}{|c|}{\sigma_1=0, \sigma_2=\frac{1}{2}}\\\hline
(\odd,\even) &
e_{61}\; e_{64}\; e_{51}\; e_{54}
&
e_{61}\; e_{65}\; e_{51}\; e_{55}
&
e_{61}\; e_{66}\; e_{51}\; e_{56}
&
e_{61}\; e_{64}\; e_{51}\; e_{54}
&
e_{64}\; e_{66}\; e_{54}\; e_{56}
&
e_{64}\; e_{65}\; e_{54}\; e_{55}
\\\hline
(\even, \odd) &
(e_{11})\; e_{14}\; e_{51}\; e_{54}
&
(e_{11})\; e_{15}\; e_{51}\; e_{55}
&
(e_{11})\; e_{16}\; e_{51}\; e_{56}
&
(e_{11})\; e_{14}\; e_{51}\; e_{54}
&
e_{14}\; e_{16}\; e_{54}\; e_{56}
&
e_{14}\; e_{15}\; e_{54}\; e_{55}
\\\hline
 (\odd,\odd) &
e_{41}\; e_{44}\; e_{51}\; e_{54}
&
e_{41}\; e_{45}\; e_{51}\; e_{55}
&
e_{41}\; e_{46}\; e_{51}\; e_{56}
&
e_{41}\; e_{44}\; e_{51}\; e_{54}
&
e_{44}\; e_{46}\; e_{54}\; e_{56}
&
e_{44}\; e_{45}\; e_{54}\; e_{55}
\\\hline\hline
\multicolumn{7}{|c|}{\sigma_1=\sigma_2=\frac{1}{2}} \\\hline
(\odd,\even) &
e_{61}\; e_{64}\; e_{51}\; e_{54}
&
e_{61}\; e_{65}\; e_{51}\; e_{55}
&
e_{61}\; e_{66}\; e_{51}\; e_{56}
&
e_{61}\; e_{64}\; e_{51}\; e_{54}
&
e_{64}\; e_{66}\; e_{54}\; e_{56}
&
e_{64}\; e_{65}\; e_{54}\; e_{55}
\\\hline
(\even, \odd) &
e_{41}\; e_{44}\; e_{61}\; e_{64}
&
e_{41}\; e_{45}\; e_{61}\; e_{65}
&
e_{41}\; e_{46}\; e_{61}\; e_{66}
&
e_{41}\; e_{44}\; e_{61}\; e_{64}
&
e_{44}\; e_{46}\; e_{64}\; e_{66}
&
e_{44}\; e_{45}\; e_{64}\; e_{65}
\\\hline
 (\odd,\odd) &
(e_{11})\; e_{14}\; e_{61}\; e_{64}
&
(e_{11})\; e_{15}\; e_{61}\; e_{65}
&
(e_{11})\; e_{16}\; e_{61}\; e_{66}
&
(e_{11})\; e_{14}\; e_{61}\; e_{64}
&
e_{14}\; e_{16}\; e_{64}\; e_{66}
&
e_{14}\; e_{15}\; e_{64}\; e_{65}
\\\hline\hline\hline
& \multicolumn{3}{|c|}{\sigma_3=0, \sigma_4=\frac{1}{2}}
& \multicolumn{3}{|c|}{\sigma_3=\sigma_4=\frac{1}{2}}\\\hline\hline
\multicolumn{7}{|c|}{\sigma_1=\sigma_2=0}\\\hline
(\odd,\even) &
e_{15}\; e_{16}\; e_{45}\; e_{46}
&
(e_{11})\; e_{15}\; e_{41}\; e_{45}
&
e_{15}\; e_{14}\; e_{44}\; e_{45}
&
e_{15}\; e_{16}\; e_{45}\; e_{46}
&
e_{14}\; e_{16}\; e_{44}\; e_{46}
&
(e_{11})\; e_{16}\; e_{41}\; e_{46}
\\\hline
(\even, \odd) &
e_{15}\; e_{16}\; e_{55}\; e_{56}
&
(e_{11})\; e_{15}\; e_{51}\; e_{55}
&
e_{14}\; e_{15}\; e_{54}\; e_{55}
&
e_{15}\; e_{16}\; e_{55}\; e_{56}
&
e_{14}\; e_{16}\; e_{54}\; e_{56}
&
(e_{11})\; e_{16}\; e_{51}\; e_{56}
\\\hline
 (\odd,\odd) &
e_{15}\; e_{16}\; e_{65}\; e_{66}
&
(e_{11})\; e_{15}\; e_{61}\; e_{65}
&
e_{14}\; e_{15}\; e_{64}\; e_{65}
&
e_{15}\; e_{16}\; e_{65}\; e_{66}
&
e_{14}\; e_{16}\; e_{64}\; e_{66}
&
(e_{11})\; e_{16}\; e_{61}\; e_{66}
\\\hline\hline
\multicolumn{7}{|c|}{\sigma_1=\frac{1}{2}, \sigma_2=0} \\\hline
(\odd,\even) &
e_{15}\; e_{16}\; e_{45}\; e_{46}
&
(e_{11})\; e_{15}\; e_{41}\; e_{45}
&
e_{14}\; e_{15}\; e_{44}\; e_{45}
&
e_{15}\; e_{16}\; e_{45}\; e_{46}
&
e_{14}\; e_{16}\; e_{44}\; e_{46}
&
(e_{11})\; e_{16}\; e_{41}\; e_{46}
\\\hline
(\even, \odd) &
e_{65}\; e_{66}\; e_{45}\; e_{46}
&
e_{61}\; e_{65}\; e_{41}\; e_{45}
&
e_{64}\; e_{65}\; e_{44}\; e_{45}
&
e_{65}\; e_{66}\; e_{45}\; e_{46}
&
e_{64}\; e_{66}\; e_{44}\; e_{46}
&
e_{61}\; e_{66}\; e_{41}\; e_{46}
\\\hline
 (\odd,\odd) &
e_{55}\; e_{56}\; e_{45}\; e_{46}
&
e_{51}\; e_{55}\; e_{41}\; e_{45}
&
e_{54}\; e_{55}\; e_{44}\; e_{45}
&
e_{55}\; e_{56}\; e_{45}\; e_{46}
&
e_{54}\; e_{56}\; e_{44}\; e_{46}
&
e_{51}\; e_{56}\; e_{41}\; e_{46}
\\\hline\hline
\multicolumn{7}{|c|}{\sigma_1=0, \sigma_2=\frac{1}{2}}\\\hline
(\odd,\even) &
e_{65}\; e_{66}\; e_{55}\; e_{56}
&
e_{61}\; e_{65}\; e_{51}\; e_{55}
&
e_{64}\; e_{65}\; e_{54}\; e_{55}
&
e_{65}\; e_{66}\; e_{55}\; e_{56}
&
e_{64}\; e_{66}\; e_{54}\; e_{56}
&
e_{61}\; e_{66}\; e_{51}\; e_{56}
\\\hline
(\even, \odd) &
e_{15}\; e_{16}\; e_{55}\; e_{56}
&
(e_{11})\; e_{15}\; e_{51}\; e_{55}
&
e_{14}\; e_{15}\; e_{54}\; e_{55}
&
e_{15}\; e_{16}\; e_{55}\; e_{56}
&
e_{14}\; e_{16}\; e_{54}\; e_{56}
&
(e_{11})\; e_{16}\; e_{51}\; e_{56}
\\\hline
 (\odd,\odd) &
e_{45}\; e_{46}\; e_{55}\; e_{56}
&
e_{41}\; e_{45}\; e_{51}\; e_{55}
&
e_{44}\; e_{45}\; e_{54}\; e_{55}
&
e_{45}\; e_{46}\; e_{55}\; e_{56}
&
e_{44}\; e_{46}\; e_{54}\; e_{56}
&
e_{41}\; e_{46}\; e_{51}\; e_{56}
\\\hline\hline
\multicolumn{7}{|c|}{\sigma_1=\sigma_2=\frac{1}{2}} \\\hline
(\odd,\even) &
e_{65}\; e_{66}\; e_{55}\; e_{56}
&
e_{61}\; e_{65}\; e_{51}\; e_{55}
&
e_{64}\; e_{65}\; e_{54}\; e_{55}
&
e_{65}\; e_{66}\; e_{55}\; e_{56}
&
e_{64}\; e_{66}\; e_{54}\; e_{56}
&
e_{61}\; e_{66}\; e_{51}\; e_{56}
\\\hline
(\even, \odd) &
e_{45}\; e_{46}\; e_{65}\; e_{66}
&
e_{41}\; e_{45}\; e_{61}\; e_{65}
&
e_{44}\; e_{45}\; e_{64}\; e_{65}
&
e_{45}\; e_{46}\; e_{65}\; e_{66}
&
e_{44}\; e_{46}\; e_{64}\; e_{66}
&
e_{41}\; e_{46}\; e_{61}\; e_{66}
\\\hline
 (\odd,\odd) &
e_{15}\; e_{16}\; e_{65}\; e_{66}
&
(e_{11})\; e_{15}\; e_{61}\; e_{65}
&
e_{14}\; e_{15}\; e_{64}\; e_{65}
&
e_{15}\; e_{16}\; e_{65}\; e_{66}
&
e_{14}\; e_{16}\; e_{64}\; e_{66}
&
(e_{11})\; e_{16}\; e_{61}\; e_{66}
\\\hline
      \end{array}
    \end{equation*}
  \end{center}
\caption{Relation between fixed points and wrapping numbers on $T^2_1 \times T^2_2$.
The fixed point $e_{11}$ does not give rise to any exceptional cycle. The bulk 2-cycle
specified by the wrapping numbers can be displaced from the origin by
$\sum_{i=1}^4 \sigma_i \pi_i$ with $\sigma_i \in \{0, 1/2\}$.}
\label{AppTab:WrapFPZ6}
\end{table}

\renewcommand{\arraystretch}{1.3}
\begin{table}[ht]
  \begin{center}
    \begin{equation*}
      \begin{array}{|l|c|} \hline
\text{2-cycle}\otimes\text{1-cycle} & \text{Exceptional 3-cycle}\\\hline\hline
e_{11} \otimes (n_3\pi_5+m_3\pi_6)  & ----   \\\hline\hline
e_{14} \otimes (n_3\pi_5+m_3\pi_6)  &  n_3\varepsilon_2+m_3\Tilde{\varepsilon}_2 \\\hline
e_{15} \otimes (n_3\pi_5+m_3\pi_6)  &  n_3(\Tilde{\varepsilon}_2 - \varepsilon_2)-m_3\varepsilon_2
 \\\hline
e_{16} \otimes (n_3\pi_5+m_3\pi_6)  & -n_3\Tilde{\varepsilon}_2-m_3(\Tilde{\varepsilon}_2
- \varepsilon_2)\\\hline\hline
e_{41} \otimes (n_3\pi_5+m_3\pi_6)  &  n_3\varepsilon_1+m_3\Tilde{\varepsilon}_1 \\\hline
e_{51} \otimes (n_3\pi_5+m_3\pi_6)  &  n_3(\Tilde{\varepsilon}_1 - \varepsilon_1)-m_3\varepsilon_1
\\\hline
e_{61} \otimes (n_3\pi_5+m_3\pi_6)  & -n_3\Tilde{\varepsilon}_1
-m_3(\Tilde{\varepsilon}_1 - \varepsilon_1)\\\hline\hline
e_{44} \otimes (n_3\pi_5+m_3\pi_6)  &  n_3\varepsilon_3+m_3\Tilde{\varepsilon}_3 \\\hline
e_{45} \otimes (n_3\pi_5+m_3\pi_6)  &  n_3\varepsilon_4+m_3\Tilde{\varepsilon}_4\\\hline
e_{46} \otimes (n_3\pi_5+m_3\pi_6)  &  n_3\varepsilon_5+m_3\Tilde{\varepsilon}_5\\\hline\hline
e_{54} \otimes (n_3\pi_5+m_3\pi_6)  &  n_3(\Tilde{\varepsilon}_5 - \varepsilon_5)-m_3\varepsilon_5
\\\hline
e_{55} \otimes (n_3\pi_5+m_3\pi_6)  &  n_3(\Tilde{\varepsilon}_3 - \varepsilon_3)-m_3\varepsilon_3
\\\hline
e_{56} \otimes (n_3\pi_5+m_3\pi_6)  &  n_3(\Tilde{\varepsilon}_4 - \varepsilon_4)-m_3\varepsilon_4
\\\hline\hline
e_{64} \otimes (n_3\pi_5+m_3\pi_6)  & -n_3\Tilde{\varepsilon}_4
-m_3(\Tilde{\varepsilon}_4 - \varepsilon_4)\\\hline
e_{65} \otimes (n_3\pi_5+m_3\pi_6)  & -n_3\Tilde{\varepsilon}_5
-m_3(\Tilde{\varepsilon}_5 - \varepsilon_5)\\\hline
e_{66} \otimes (n_3\pi_5+m_3\pi_6)  & -n_3\Tilde{\varepsilon}_3
-m_3(\Tilde{\varepsilon}_3 - \varepsilon_3)\\\hline
      \end{array}
    \end{equation*}
  \end{center}
\caption{Relation between orbits of fixed points and cycles as read off from~(\ref{Eq:exceptionalcycles}).}
\label{AppTab:FPandCycles}
\end{table}



\section{Some loop an tree channel results}\label{AppSec:Amplitude}

\subsection{Tree channel bulk part}\label{AppSubsec:BulkBoundary}

The oscillator expansion of an untwisted boundary state~(\ref{Eq:UntwistedBounState})
with spin structure $\eta=\pm 1$ and  relative angles
$\pi\varphi^k_a$ w.r.t. $\pi_{2k-1}$ is given by~\cite{Blumenhagen:2000wh,Forste:2001gb}
\begin{multline}
|D6;(n^a_i,m^a_i),\eta \rangle\\
\sim \exp\left\{-\sum_{k=0}^3\sum_{n} \frac{e^{2\pi i
      \varphi^k_a}}{n}\alpha_{-n}^k  \Tilde{\alpha}_{-n}^k
-i\eta \sum_{k=0}^3 \sum_{r} e^{2\pi i \varphi^k_a} \psi_{-r}^k \Tilde{\psi}_{-r}^k
+h.c.
\right\}|0,\eta;p^k,w^k;\tau_i,\sigma_i\rangle.
\end{multline}
In order to shorten the notation, the non-compact coordinates are denoted by $k=0$ and the
corresponding angle is $\varphi^0_a \equiv 0$.

In addition to the Kaluza Klein momenta and windings existing
in toroidal compactifications,
for fractional branes
discrete Wilson lines on $T^2_1 \times T^2_2$ parameterized by
$\tau_1, \tau_2$  arise from the $\Z_2$ fixed
points~\cite{Blumenhagen:2002gw}, see eq.~(\ref{Eq:FixpointsWilsonlines}).
A bulk brane can be displaced from the origin by $\sum_{i=1}^6 \sigma_i \pi_i$,
with arbitrary values $\sigma_i \in [0,1)$,
whereas for a fractional brane the displacement is discretized on two tori,
$\sigma_i \in \{0,1/2\}$ for $i=1,\ldots,4$.

Relative Wilson lines between two different boundary states which
are parallel on a 2-torus give rise to complex phases at each mass
level, i.e. if $L$ is the dimensionless length of the 1-cycle
wrapped by the two branes, $R^2$ the volume of the corresponding
2-torus, $\frac{\tau}{2} \in \{0,1/2\}$ the relative Wilson line
and $\sigma$ the spatial separation, the tree channel zero mode
contributions are given by
\begin{equation}\label{Eq:treelatticecontr}
\Tilde{\cal L}(l)=\left( \sum_{r \in \Z} e^{-\pi l r^2 L^2R^2/\alpha'+\pi i r \tau}\right)
\left( \sum_{s \in \Z} e^{-\pi l s^2 L^2\alpha'/R^4 + 2\pi i s \sigma }\right).
\end{equation}
Modular transformation leads to the loop channel annulus contribution
\begin{equation}\label{Eq:looplatticecontr}
{\cal L}(t)=\left( \sum_{r \in \Z} e^{-2\pi t (r-\tau/2)^2 \alpha'/(R^2L^2)}\right)
\left( \sum_{s \in \Z} e^{-2 \pi t (s-\sigma)^2 R^4/(L^2\alpha')}\right),
\end{equation}
which confirms that strings stretching between parallel branes with
relative Wilson lines and/or spatial separation do not carry
massless modes.

\subsection{Tree channel twisted part}\label{Subsubsec:BoundaryTwisted}

The oscillator modding of a twisted boundary state~(\ref{Eq:TwistedBounState})
is shifted by the twist vector
$v=(1/2,-1/2,0)$,
\begin{equation}
\begin{aligned}
|D6; (n^a_3,m^a_3),e_{ij}\eta\rangle
& \sim \exp\Bigl\{-\sum_{k=0,3}\sum_{n} \frac{e^{2\pi i
    \varphi^k_a}}{n}\alpha_{-n}^k \Tilde{\alpha}_{-n}^k
-i\eta \sum_{k=0,3} \sum_{r} e^{2\pi i \varphi^k_a} \psi_{-r}^k \Tilde{\psi}_{-r}^k \\
-\sum_{j=1,2}\sum_{n} \frac{e^{2\pi i \varphi^j_a}}{n} &\alpha_{-n+v_j}^j \Tilde{\alpha}_{-n+v_j}^j
-i\eta \sum_{j=1,2} \sum_{r} e^{2\pi i \varphi^j_a} \psi_{-r+v_j}^j \Tilde{\psi}_{-r+v_j}^j
+h.c. \Bigr\}
|0,\eta;p^3,w^3,e_{ij}\rangle,
\end{aligned}
\end{equation}
and the discrete Wilson lines enter the boundary states only as relative signs $\alpha_{ij}$
between the twisted sector contributions, compare eq.~(\ref{Eq:TwistedBounState}).

The crosscap states do not have any twisted contributions.


\subsection{Oscillator contributions to the amplitudes}\label{AppSubsec:OsciCont}

The tree channel oscillator contributions to the annulus and M\"obius strip amplitude are of the form
\begin{equation}\label{Eq:OscTreeAM}
\begin{aligned}
  \Tilde{\cal A}^{\alpha \beta}_{v,(\varphi_1,\varphi_2,\varphi_3)} &=
  \frac{\vartheta \left[\begin{array}{c} \alpha \\ \beta \end{array} \right]}{\eta^3}
  \prod_{i=1}^3
  \frac{\vartheta \left[\begin{array}{c} \alpha-v_i \\ \varphi_i +\beta \end{array} \right]}
  {\vartheta \left[\begin{array}{c} 1/2-v_i \\  \varphi_i +1/2 \end{array} \right]}(2l)\\
  \Tilde{\cal M}^{\alpha \beta}_{(\varphi_1,\varphi_2,\varphi_3)} &=
  \frac{\vartheta \left[\begin{array}{c} \alpha \\ \beta \end{array} \right]}{\eta^3}
  \prod_{i=1}^3
  \frac{\vartheta \left[\begin{array}{c} \alpha \\ \varphi_i +\beta \end{array} \right]}
  {\vartheta \left[\begin{array}{c} 1/2 \\  \varphi_i +1/2 \end{array} \right]}(2l-i/2)
\end{aligned}
\end{equation}
where $v=0$ for the untwisted and $v=(1/2,-1/2,0)$ for the $\Z_2$ twisted annulus contributions.
$\alpha=0,1/2$ corresponds to NSNS and RR contributions, respectively and $\beta=0,1/2$ labels
contributions from identical and opposite spin structures.
For more details on the notation see e.g. the appendices of~\cite{Blumenhagen:1999ev,Forste:2000hx}.

For each vanishing angle $\varphi_i=0$ and vanishing twist $v_i$, the corresponding denominator
has to be replaced,
$\vartheta \left[\begin{array}{c} 1/2 \\  \varphi_i +1/2 \end{array} \right]
\stackrel{\varphi_i \rightarrow 0}{\longrightarrow} \eta^3$.

The loop channel oscillator contributions are given by
\begin{equation}\label{Eq:OscLoopAM}
\begin{aligned}
{\cal A}^{A,B}_{v,(\varphi_1,\varphi_2,\varphi_3)} &=i
\frac{\vartheta \left[\begin{array}{c} A \\ B \end{array} \right]}{\eta^3}
  \prod_{i=1}^3
  \frac{\vartheta \left[\begin{array}{c}  A-\varphi_i \\ B-v_i  \end{array} \right]}
  {\vartheta \left[\begin{array}{c}  1/2-\varphi_i \\ 1/2-v_i  \end{array} \right]}(t)\\
{\cal M}^{A,B}_{(\varphi_1,\varphi_2,\varphi_3)}
&= i  \frac{\vartheta \left[\begin{array}{c} A \\ B \end{array} \right]}{\eta^3}
\prod_{i=1}^3
 \frac{\vartheta \left[\begin{array}{c} A +2 \varphi_i\\ B -\varphi_i \end{array} \right]}
{\vartheta \left[\begin{array}{c} 1/2 +2 \varphi_i \\ 1/2- \varphi_i \end{array}\right]}(t-i/2)
\end{aligned}
\end{equation}
where $v=0$ corresponds to an open string state without  and $v=(1/2,-1/2,0)$ with $\Z_2$ insertion.
The modification for vanishing angle is identical to the tree channel result up to a factor of $i$.

The modular transformation $l=1/(\kappa t)$ (where $\kappa=2$ or $8$ for the
annulus or M\"obius strip, respectively) for three non-vanishing
angles is given by
\begin{equation}\label{Eq:ModularTrafoAM}
\begin{aligned}
 \Tilde{\cal A}^{\alpha \beta}_{v,(\varphi_1,\varphi_2,\varphi_3)}
&=\frac{e^{\pi i (2 \alpha -1) \sum_{i=1}^3 \varphi_i}}{t}
{\cal A}^{\beta,\alpha}_{v,(\varphi_1,\varphi_2,\varphi_3)},\\
 \Tilde{\cal M}^{\alpha \beta}_{(\varphi_1,\varphi_2,\varphi_3)}
&=\frac{e^{2\pi i \alpha}}{2} \frac{e^{4(\alpha+\beta) \pi i \sum_{i=1}^3 \varphi_i}}{t}
{\cal M}^{\alpha,1/2-(\alpha+\beta)}_{(\varphi_1,\varphi_2,\varphi_3)}.
\end{aligned}
\end{equation}
Each vanishing angle modifies these equations by a factor of $1/t$ for the
annulus and $1/(2t)$ for the M\"obius  strip.


\section{Some results for $T^6/(\Z_6 \times \Z_3)$}\label{AppSec:Z6Z3}

The orbifold generators $\Theta$ and $\omega$ are represented by the two shift
vectors $v=(1/6,-1/6,0)$ and
$w=(0,1/3,-1/3)$, respectively. The Hodge numbers are given by (see e.g.~\cite{Klein:2000qw}
and also \cite{Forste:2000hx} for the closed string spectrum)
\begin{equation}
\begin{aligned}
h^U_{1,1}&=3, \qquad h^{\Theta^3}_{1,1}=4, \qquad h^{fixplanes-not-\Z_2}_{1,1}=36,
\qquad h^{fixpoints}_{1,1}=30,\\
h^U_{2,1}&=0, \qquad h^{\Theta^3}_{2,1}=1, \qquad h^{fixplanes-not-\Z_2}_{2,1}=0,
\qquad h^{fixpoints}_{2,1}=0.
\end{aligned}
\end{equation}
The allowed compactification lattices are as depicted in
figure~\ref{Fig:Z6torifixedpoints}, and the
fundamental bulk 3-cycles can be chosen to be identical (up to normalization)  to
those displayed in eq.~(\ref{Eq:bulkcycles}).\\
The orbits of wrapping numbers which describe all possible factorizable 3-cycles are given by
\begin{equation}
\small
\begin{aligned}
& \left( \begin{array}{cc}
n_1 & m_1 \\ n_2 & m_2 \\n_3 & m_3
\end{array}\right)\stackrel{\Theta}{\longrightarrow}
\left(\begin{array}{cc}
-m_1 & n_1+m_1 \\ (n_2+m_2) & -n_2 \\ n_3 & m_3
\end{array}\right)\stackrel{\Theta}{\longrightarrow}
\left(\begin{array}{cc}
-(n_1+m_1) & n_1 \\ m_2 & -(n_2+m_2) \\ n_3 & m_3
\end{array}\right)\\
&\downarrow \omega\\
& \left( \begin{array}{cc}
n_1 & m_1 \\ -(n_2+m_2) & n_2 \\m_3 & -(n_3+m_3)
\end{array}\right)\stackrel{\Theta}{\longrightarrow}
\left(\begin{array}{cc}
-m_1 & n_1+m_1 \\ -m_2 & (n_2+m_2) \\ m_3 & -(n_3+m_3)
\end{array}\right)\stackrel{\Theta}{\longrightarrow}
\left(\begin{array}{cc}
-(n_1+m_1) & n_1 \\ n_2 & m_2 \\ m_3 & -(n_3+m_3)
\end{array}\right)\\
&\downarrow \omega\\
& \left( \begin{array}{cc}
n_1 & m_1 \\ m_2 & -(n_2+m_2) \\-(n_3+m_3) & n_3
\end{array}\right)\stackrel{\Theta}{\longrightarrow}
\left(\begin{array}{cc}
-m_1 & n_1+m_1 \\ -n_2 & -m_2 \\ -(n_3+m_3) & n_3
\end{array}\right)\stackrel{\Theta}{\longrightarrow}
\left(\begin{array}{cc}
-(n_1+m_1) & n_1 \\ -(n_2+m_2) & n_2 \\ -(n_3+m_3) & n_3
\end{array}\right)
\end{aligned}
\end{equation}
and lead (up to normalization) to the coefficients~(\ref{Eq:bulkCoeffYZ}) computed for the
$T^6/\Z_6$ case.\\
Only two exceptional 3-cycles arise. The fixed points on $T^2_1$ are permuted under $\Theta$ as
in~(\ref{Eq:thetaonfixedpoints}) and are fixed under $\omega$. On $T^2_2$, the
permutations are given by
\begin{equation}
\Theta (4)  =\omega (4)=  6, \qquad \Theta (5)=\omega (5)= 4, \qquad \Theta (6)=\omega (6)= 5.
\end{equation}
On $T^2_3$, only $\omega$ fixed points occur, which transform trivially under $\Theta$ and do not
contribute to exceptional 3-cycles. Two linearly independent, orbifold invariant exceptional 3-cycles
with vanishing self-intersection
can be expressed in terms of~(\ref{Eq:exceptionalcycles}),
\begin{equation}
\begin{aligned}
\zeta_1 = & (1+\Theta+\Theta^2)(1+\omega+\omega^2)e_{44}\otimes\pi_5\\
\sim & \varepsilon_3-\varepsilon_4+\Tilde{\varepsilon}_4-\Tilde{\varepsilon}_5,
\\
\zeta_2 = & (1+\Theta+\Theta^2)(1+\omega+\omega^2)e_{55}\otimes\pi_5\\
\sim & -\varepsilon_3+\varepsilon_5+\Tilde{\varepsilon}_3-\Tilde{\varepsilon}_4.
\end{aligned}
\end{equation}

\end{appendix}


\end{document}